\documentclass[12pt]{article}

\usepackage{amscd, amsmath}
\usepackage{amsthm, amssymb}
\usepackage{array}

\newcommand{\CC}{\mathbb{C}}
\newcommand{\RR}{\mathbb{R}}
\newcommand{\BB}{\mathcal B}
\newcommand{\AAA}{\mathcal A}
\newcommand{\PP}{\mathcal P}
\newcommand{\HH}{\mathcal H}
\newcommand{\he}{\hat{e}}

\newcommand{\db}{\bar{\partial}}
\newcommand{\dd}{{\rm d}}
\newcommand{\MM}{{\mathcal M}}
\newcommand{\g}{\mathfrak g}

\newcommand{\GG}{\mathcal G}

\newcommand{\LL}{\mathcal L}
\newcommand{\VVV}{\mathcal V}

\newcommand{\dA}{\dot{A}}
\newcommand{\dphi}{\dot{\phi}}
\newcommand{\ev}{ev}
\newcommand{\balpha}{{\bar{\alpha}}}
\newcommand{\bbeta}{{\bar{\beta}}}
\newcommand{\bk}{{\bar{k}}}
\newcommand{\bz}{{\bar{z}}}

\newtheorem{theorem}{Theorem}[section]
\newtheorem{prop}[theorem]{Proposition}
\newtheorem{cor}[theorem]{Corollary}

\newtheorem{assumption (i)}[theorem]{Assumption (i)}
\newtheorem{assumption (ii)}[theorem]{Assumption (ii)}
\newtheorem{conjecture}[theorem]{Conjecture}

\theoremstyle{remark}
\newtheorem*{rem}{{\bf Remark}}

\theoremstyle{remark}

\theoremstyle{remark}

\newenvironment*{prooff}{\noindent {\bf Proof.}}{\hfill $\qed$ \vspace{.3cm}}
\newenvironment*{prooff1.1}{\noindent {\bf Proof of theorem \ref{thm1.1}.}}{\hfill $\qed$ \vspace{.3cm}}

\begin{document}

\begin{titlepage}
\title{
\vskip -70pt
\begin{flushright}
{\normalsize ITFA-2010-09 \ }\\
\end{flushright}
\vskip 45pt
{\bf On the $L^2$-metric of vortex moduli spaces}
}
\vspace{3cm}

\author{{J. M. Baptista} \thanks{ e-mail address:
    j.m.baptista@uva.nl}  \\
{\normalsize {\sl Institute for Theoretical Physics} \thanks{ address: Valckenierstraat 65, 1018 XE Amsterdam, The Netherlands}
} \\
{\normalsize {\sl University of Amsterdam}} 
}

\date{March 2010}

\maketitle

\thispagestyle{empty}
\vspace{1.5cm}
\vskip 20pt
{\centerline{{\large \bf{Abstract}}}}
\vspace{.2cm}
We derive general expressions for the K\"ahler form of the $L^2$-metric in terms of standard 2-forms on vortex moduli spaces. In the case of abelian vortices in gauged linear sigma-models, this allows us to compute explicitly the K\"ahler class of the $L^2$-metric.
As an application we compute the total volume of the moduli space of abelian semi-local vortices. In the strong coupling limit, this then leads to conjectural formulae for the volume of the space of holomorphic maps from a compact Riemann surface to projective space. Finally we show that the localization results of Samols in the abelian Higgs model extend to more general models. These include linear non-abelian vortices and vortices in gauged toric sigma-models.

\vspace{.45cm}

\end{titlepage}

\tableofcontents

\section{Introduction}

Vortices are the natural solitons of gauged sigma-models at critical coupling. Within each topological sector they minimize the Yang-Mills-Higgs energy functional
\begin{equation}
E(A, \phi) \ = \ \int_{\Sigma} \ \frac{1}{2\, e^2} \: |F_A|^2 \ + \ |\dd^A \phi|^2 \ + \ 2\, e^2 \: |\mu \circ \phi |^2 \ .
\label{1.0}
\end{equation}
To construct this functional and define the model we need two K\"ahler manifolds $\Sigma$ and $X$, and a hamiltonian action of  a Lie group $G$ on $X$. The fields of the theory are then a connection $A$ on a principal $G$-bundle $P \rightarrow \Sigma$ and a section $\phi$ of the associated bundle $P\times_G X \rightarrow \Sigma$. This section $\phi$ can be locally regarded as a map $\Sigma \rightarrow X$, so one can compose it with the moment map $\mu: X \rightarrow {\rm Lie}\; G$ and take the squared norm using an Ad-invariant inner product on the Lie algebra. The resulting function $| \mu \circ \phi |^2$ is then used as the potential at critical coupling in the energy $(\ref{1.0})$. The remaining two terms in the functional are the usual Maxwell term -- the squared norm of the curvature of $A$ -- and a standard covariant derivative term.

The well-known Bogomolny argument can be used to show that the energy (\ref{1.0}) is minimized by the field configurations $(A , \phi)$ that satisfy 
\begin{align}
&\bar{\partial}^A \phi \ = \ 0    \label{1.2} \\
&\Lambda F_A  \: +\:  2  e^2\;  \mu \circ \phi \ = \ 0 \ \nonumber \\
& F_A^{0,2} \ = \ 0 \ , \nonumber
\end{align}
which are usually called the vortex equations. Here the symbol $\Lambda F_A$ represents the inner product $\langle \omega_\Sigma , F_A \rangle$ of the curvature with the K\"ahler form of $\Sigma$. In the special and important case when $\Sigma$ is a Riemann surface, the last equation is trivially satisfied and the vortex equations reduce to 
\begin{align}
&\bar{\partial}^A \phi \ = \ 0    \label{1.3} \\
&\ast F_A  \: +\:  2  e^2\;  \mu \circ \phi \ = \ 0 \ ,\nonumber
\end{align}
where $\ast$ is the Hodge star operator on $\Sigma$. This general version of the vortex equations was first found in $\cite{Mundet, C-G-S}$ and extends to arbitrary hamiltonian targets $X$ the previously known versions of the equations, which had been written mostly for linear targets acted upon by linear $G$-representations. For example the classical abelian Higgs model corresponds to the choices $G = U(1)$ and $X=\CC$. These general vortex equations can be used to define the so-called Hamiltonian (or gauged) Gromov-Witten invariants of $X$ \cite{C-G-M-S}. From a physics perspective, the vortex solutions in two dimensions are the BPS configurations of a ${\mathcal N}=2$ supersymetric gauged sigma-model \cite{Baptista2}.

$\ $

When studying the vortex equations one of the main objects of interest is the moduli space $\MM$ of solutions, i.e. the set of vortex solutions quotiented by the action of the group of $G$-gauge transformations. In general this is a complicated space that is very hard to describe, but it is also a space with many interesting features. One of the most interesting properties of $\MM$ is that it possesses natural homology and cohomology classes, and so by taking cup products and intersections one can obtain natural intersection numbers associated to $\MM$, which is exactly how the Hamiltonian Gromov-Witten invariants are defined \cite{C-G-M-S, Baptista1}. Another important property of $\MM$ is that is possesses a natural K\"ahler structure (see e.g. \cite{Mundet}), with Riemannian metric defined by 
\begin{equation}
g_\MM (\dot{A}_1 + \dot{\phi}_1 , \dot{A}_2 + \dot{\phi}_2  ) \ = \ \int_\Sigma \ \frac{1}{4e^2} \: k_{ab} \: \dot{A}_1^a \wedge  \ast_{\Sigma} \, \dot{A}_2^b  \ + \  g_X ( \dot{\phi}_1 , \dot{\phi}_2 )\ {\rm vol}_\Sigma
\label{1.1}
\end{equation}
and a compatible complex structure defined by 
\begin{equation}
J_\MM \: (\dA ,  \dphi) \ = \ (\: -i \dA_{\alpha}\, \dd z^\alpha \: + \: i \dA_{\bar{\alpha}}\, \dd \bz^{\bar{\alpha}}\: ,\: J_X \: \dphi\:  ) \ ,
\label{1.4}
\end{equation}
where $\{z^\alpha \}$ and $\ast_\Sigma$ are, respectively, a set of complex coordinates and the Hodge operator on $\Sigma$.
Observe that (\ref{1.1}) in fact defines a metric in the space $\VVV$ of all vortex solutions $(A, \phi)$, but since the tangent space $T_{[A, \phi]} \MM$ can be identified with the subspace of $T_{(A, \phi)} \VVV$ that is orthogonal to gauge transformations, i.e. orthogonal to the kernel of the natural projection $T_{(A, \phi)} \VVV \rightarrow T_{[A, \phi]} \MM$, the same expression (\ref{1.1}) also determines a metric on $\MM$. Besides being geometrically very natural, this $L^2$-metric on the moduli space is also physically important, essentially because the induced norm $|\!|\,(\dA , \dphi)\, |\!|_{\MM}^2$can be interpreted as the kinetic energy of a dynamical vortex solution that evolves slowly with time. This fact was firstly recognized by Manton for the case of monopoles on $\RR^3$ \cite{Manton1}, and constitutes the basis of an approximation method that has been widely used to describe the low-energy dynamics of various types on solitons \cite{Manton-Sutcliffe}.

In the case of vortices, though, the problem of describing the metric $g_\MM$ has proved to be a difficult one, essentially because no non-trivial vortex solutions are known explicitly. This study is more developed in the simplest case of vortices with $G= U(1)$ and $X=\CC$, i.e. in the abelian Higgs model, where some of the most remarkable results are Samols' formula for $g_\MM$ in terms of the local behaviour of $|\phi|^2$ near its zeros \cite{Samols}, the computation of the cohomology class $[\omega_\MM]$ of the vortex K\"ahler form \cite{Manton2, Perutz}, and an asymptotic formula for $g_\MM$ in the region of the moduli space where the vortices are well-separated \cite{Manton-Speight}. In this article we basically extend the first two results beyond the abelian Higgs model, i.e. to vortices with different groups $G$ and targets $X$.
On the one hand we compute the K\"ahler class $[\omega_\MM]$ for models where $G$ is a torus acting linearly on the target $X=\CC^n$;  on the other hand we show that Samols' localization happens whenever the complexified $G^\CC$-action is free and transitive on an open dense subset of the target $X$. A more detailed description of the content of the article follows below.
Other recent results on vortex metrics include \cite{amalgama}.

$\ $

In section 2 we write down a formula for the K\"ahler form $\omega_\MM$ of the $L^2$-metric in terms of  standard 2-forms on the moduli space, namely in terms of the curvature $F_\AAA$ of the universal connection and the pull-back by the evaluation map of the equivariant K\"ahler form $\omega^G_X$ of the target $X$. When the target is a complex vector space this formula simplifies considerably, and we can express $\omega_\MM$ solely in terms of the curvature $F_{\AAA}$. This simplified formula is obtained in section 3, and generalizes a formula derived by Perutz for the abelian Higgs model \cite{Perutz}. In section 4 we apply the cohomological version of this formula to the case of abelian linear sigma-models, i.e. to the case where $G$ is a torus acting with weights $Q^a_j$ on a vector space $X =\CC^n$. Since in this abelian case there exists a rather explicit description of $\MM$ \cite{Morrison-Plesser, Werheim}, this allows us to deduce an equally explicit formula for $[\omega_\MM]$, a formula that generalizes the result of Manton and Nasir for the abelian Higgs model \cite{Manton2}. As an application, in section 5 we compute the total volume and the Einstein-Hilbert action of $(\MM , g_\MM)$ in the case of abelian semi-local vortices, i.e. in the case where $G=U(1)$ acts by scalar multiplication on $X=\CC^n$. Taking the large coupling limit $e^2 \rightarrow \infty$, this leads directly to conjectural formulae for the total volume and Einstein-Hilbert action of the space of holomorphic maps $\Sigma \rightarrow \CC {\mathbb P}^{n-1}$ equipped with its natural $L^2$-metric. This limit is explained in section 5.2. Finally, in the last two sections of the paper (which can be read independently) we explain how the localization phenomena described by Samols in the abelian Higgs model extend to the more general cases where the complexified group $G^\CC$ acts freely and transitively on an open dense subset $X^o \subset X$ of the target. This extension of Samols' localization is illustrated more concretely in two different examples: the case of linear non-abelian vortices with group G=U(N) acting on the space of complex square matrices $X = M_{N\times N}$; and the case of abelian vortices where a torus $G=T^k$ acts on a toric manifold of complex dimension $k$.

$\ $

\noindent
{\it Note:} A week before this article was completed, a paper that contains some overlap with the material of section 7.1 in the case $\Sigma =\CC$ appeared on the arXiv \cite{F-M-N-O-S}. This reference also addresses the question of finding the asymptotic form of $g_\MM$ for well-separated, linear non-abelian vortices, an interesting subject that is not discussed here.

\section{The vortex K\"ahler form}

\subsection{Universal bundle and connection}

This is essentially a preparatory section. We recall the well-known concept of universal bundle with its universal connection $\AAA$, and state the definition  of the natural 2-forms $F_\AAA$ and $ev^\ast \omega^G_X$ on the cartesian product $\MM \times \Sigma$.

Let $\VVV$ be the space of solutions of the vortex equations and let $\GG$ be the group of gauge transformations. We assume that there exists an open subset $\VVV^\ast \subset \VVV$ of irreducible solutions where the gauge transformations act freely and the quotient $\VVV^\ast / \GG \: =: \MM_{\rm smooth}$ is a finite-dimensional smooth manifold. The quotient $\MM_{\rm smooth}$ can be regarded as the smooth part of the moduli space $\MM$, and in some special instances -- e.g. the cases treated in sections 5 and 6 -- it can actually be shown that $\MM_{\rm smooth} = \MM$.

The right-action of $\GG$ on the space of vortex solutions induces a linear map from the Lie algebra of $\GG$ to the tangent spaces of $\VVV$. These derivative maps correspond to infinitesimal gauge transformations and are explicitly given by
\begin{align}
C_{(A, \phi)} \ : \  &\Omega^0 (\Sigma ;  P \times_{Ad} \g) \ \longrightarrow \ T_{(A, \phi)} \VVV     \label{2.1.0}    \\
  & \quad \epsilon = \epsilon^a e_a \  \longmapsto \  \nabla^A \epsilon \ - \ \epsilon^a \hat{e}_a  \ ,     \nonumber
\end{align}
where $\{ e_a \}$ is a basis of the Lie algebra $\g$, the $\hat{e}_a$'s are the vector fields on $X$ induced by $e_a$ and the left $G$-action, and $\nabla^A$ is the covariant derivative induced by $A$ on the adjoint bundle $P \times_{Ad} \g  \rightarrow \Sigma$. The image of this map in $T_{(A, \phi)} \VVV$ corresponds to the tangent vectors parallel to gauge transformations, or in other words to the kernel of the natural projection $T_{(A , \phi)} \VVV   \rightarrow T_{[A, \phi]} \MM$.

Now, besides acting on the vortex solutions, the group $\GG$ of gauge transformations also acts on the total space of the principal bundle $P \rightarrow \Sigma$, so one can define the smooth quotient
\begin{equation}
\PP \ = \ (\VVV^\ast \times P ) \ / \ \GG \ .
\label{2.1.1}
\end{equation}
This finite-dimensional space has a natural free $G$-action induced by the $G$-action on $P$, and so can be regarded as a principal bundle
\begin{equation}
G \hookrightarrow \  \PP \longrightarrow \ \MM_{\rm smooth} \times \Sigma \ .
\end{equation}
It is usually called the universal (or Poincar\'e) bundle. An important property of $\PP$ is that its restriction to $\Sigma$ coincides with $P$, or more precisely that for any chosen point $[A , \phi] \in \MM_{\rm smooth}$ there exists an isomorphism of $G$-bundles
\begin{equation}
\PP \ |_{[A, \phi] \times \Sigma} \ \simeq \ P \ \longrightarrow \Sigma \ .
\end{equation} 
The particular isomorphism depends of course on the initial choice of $[A , \phi]$.

Another important property of $\PP$ is that it comes equipped with a natural $G$-connection $\AAA$, sometimes called the universal connection. The curvature $F_\AAA$ of the universal connection can be regarded locally as a 2-form on $\MM_{\rm smooth} \times \Sigma$ with values on the Lie algebra $\g$. It is determined by the formulae \cite{D-K, Baptista2}
\begin{align}
F_\AAA (v_1 , v_2)\ |_{(A, \phi ,x)} \ &= \ F_A (v_1 , v_2) \ |_x     \label{2.1.2} \\ 
F_\AAA (\dA + \dphi , v) \ |_{(A, \phi , x)} \ &= \ \dA (v) \ |_x     \nonumber \\
F_\AAA (\dA_1 + \dphi_1 , \dA_2 + \phi_2 ) \ |_{(A, \phi , x)} \ &= \ [C_{(A, \phi)}^\dagger C_{(A ,\phi)} ]^{-1} (\psi) \ |_x  \ ,  \nonumber
\end{align}
where the $v$'s are tangent vectors in $T_x \Sigma$; the $\dA + \dphi$ are tangent vectors at $(A, \phi)$ that are orthogonal to gauge transformations, and hence represent a vector in $T_{[A, \phi]} \MM$; and $\psi$ is the section of the adjoint bundle $P\times_{Ad}\g  \rightarrow \Sigma$ defined by the formula
\begin{equation}
\psi \ = \ - \Big\{ 4 e^2\, k^{ab}\, (g_X)_{ts}\, \dphi^r_1\, \dphi^s_2 \, (\nabla \he_b)^t_r \ + \ 2 \, (g_\Sigma)^{\mu \nu}\,  [(\dA_1)_\nu , (\dA_2)_\mu ]  \Big\} \ e_a \ .
\label{2.1.3}
\end{equation}
Finally, a third important property of $\PP$ is that it also comes equipped with a natural $G$-equivariant map
\begin{equation}
ev \ : \ \PP \longrightarrow \ X \ ,  \qquad [A, \phi , p] \ \longmapsto \ \phi (p) \ ,
\label{2.1.4}
\end{equation}
called the evaluation map. (Here we are regarding $\phi$ as a $G$-equivariant map $P\rightarrow X$, which is the same thing as a section of $P \times_G X \rightarrow \Sigma$.) This evaluation map can be used to pull-back $G$-equivariant differential forms 
\begin{equation}
ev^\ast \ : \ \Omega^\bullet_G (X) \ \longrightarrow \ \Omega^\bullet_G (\PP) \ \simeq \ \Omega^\bullet (\MM_{\rm smooth} \times \Sigma ) \ , 
\label{2.1.5}
\end{equation}
where the last identification is just the usual Weil homomorphism defined by the connection $\AAA$ \cite{B-G-V, C-M-R, Baptista2}. This pull-back by the evaluation map can in particular be applied to the $G$-equivariant K\"ahler form of $X$
\begin{equation}
\omega_X^G \ = \ \omega_X \ - \ \xi^a \: \mu_a \ \qquad \in \  \Omega^2_G (X)
\label{2.1.6}
\end{equation}
to obtain a standard 2-form $ev^\ast \omega^G_X$ on the cartesian product $\MM_{\rm smooth} \times \Sigma$.

\subsection{The vortex K\"ahler form}

The aim of this short section is to write down an expression for the K\"ahler form $\omega_\MM$ of the vortex metric in terms of the natural 2-forms $F_\AAA$ and $ev^\ast \omega^G_X$ described above. The latter 2-forms are very standard in gauge theory, and their cohomology classes are used to define the so-called Hamiltonian (or gauged) Gromov-Witten invariants. The existence of such a formula for $\omega_\MM$ shows that the K\"ahler class $[\omega_\MM]$ can be expressed in terms of these standard cohomology classes of $\MM$, and hence that, at least in principle, quantities such as the total volume of the moduli space can be expressed in terms of Hamiltonian Gromov-Witten invariants.

\begin{theorem}
Over the smooth region $  \MM_{\rm smooth} \subset  \MM$ of the vortex moduli space the K\"ahler form of the metric $g_{\MM}$ is given by the formula
\begin{equation}
\omega_{\MM} \ = \ \int_\Sigma (\ev^\ast \omega_X^G ) \wedge \frac{\omega_\Sigma^m}{m!} \ - \ \frac{1}{4e^2}\, k_{ab}\, F_\AAA^a \wedge F^b_\AAA \wedge \frac{\omega_\Sigma^{m-1}}{(m-1)!} \ ,
\label{2.1}
\end{equation} 
where $m$ is the complex dimension of $\Sigma$ and $\omega^G_X$ and $F_\AAA$ are as in section 2.1.
\label{thm2.1}
\end{theorem}

\vspace{.2cm}
\begin{rem}
 The integrand in the formula above is a $2(m+1)$-form over the product $\MM \times \Sigma$. Applying to it fibrewise integration over $\Sigma$ yields a 2-form over $\MM$, as desired.
\end{rem}
\vspace{.2cm}

\begin{prooff}
Using the definitions of $\omega_X^G$ and $\ev^\ast$ of section 2.1 we have that
\begin{equation}
(\ev^\ast \omega_X^G ) \wedge \omega_\Sigma^m \ = \ (\ev^\ast \omega_X) \wedge \omega_\Sigma^m \ - \ (\mu_a \circ \phi)\, F_\AAA^a \wedge \omega_\Sigma^m \ .
\label{2.2}
\end{equation}  
Using the second vortex equation and standard K\"ahler geometry the second term above can be written as
\begin{align}
- \ (\mu_a \circ \phi)\,  F_\AAA^a \wedge \omega_\Sigma^m \ &= \ \frac{1}{2e^2}\, k_{ab}\, (\Lambda F_A)^b F_\AAA^a \wedge \omega_\Sigma^m    \label{2.3}  \\
&= \frac{m}{2e^2}\, k_{ab}\,  F_\AAA^a \wedge F_A^b \wedge \omega_\Sigma^{m-1} \ ,   \nonumber
\end{align}
where $k$ is an Ad-invariant product on $\g$. But looking at the different components of the definition (\ref{2.1.2}) of $F_\AAA$, we see that for the purpose of integration over $\Sigma$,
\begin{equation}
\int_\Sigma k_{ab}\,  F_\AAA^a \wedge F_\AAA^b \wedge \omega_\Sigma^{m-1} \ = \ \int_\Sigma k_{ab}\, (2  F_\AAA^a \wedge F_A^b  \ +\  \psi_\mu^a \wedge \dd x^\mu \wedge \psi_\nu^b \wedge \dd x^\nu ) \wedge \omega_\Sigma^{m-1} \ ,
\label{2.5}
\end{equation} 
where $\{ x^\mu \}$ are coordinates on $\Sigma$ and 
by definition $(\psi_\mu^a \wedge \dd x^\mu) (\dA + \dphi , v) =  \dA^a_\mu v^\mu$. This is because a form over $\MM \times \Sigma$ contributes to the integral only if it has degree $2m$ on the $\Sigma$-side. Bringing everything together we see that the form $\omega_\MM$ defined by (\ref{2.1}) can be written as 
\begin{equation}
\omega_\MM \ = \ \int_\Sigma (\ev^\ast \omega_X ) \wedge \frac{\omega_\Sigma^m}{m!} \ + \ \frac{1}{4e^2}\, k_{ab}\, \psi_\mu^a \wedge \psi_\nu^b \wedge \dd x^\mu \wedge \dd x^\nu  \wedge \frac{\omega_\Sigma^{m-1}}{(m-1)!} \ . 
\label{2.6}
\end{equation}
This means that, when contracted with tangent vectors $(\dA + \dphi )$ orthogonal to gauge transformations, $\omega_\MM$ satisfies
\begin{equation}
\omega_\MM (\dA_1 + \dphi_1 , \dA_2 + \dphi_2) \ = \ \int_\Sigma \omega_X (\dphi_1 , \dphi_2)\, \frac{\omega_\Sigma^m}{m!} \ +\ \frac{1}{4e^2}\, k_{ab} \, \dA_1^a \wedge \dA_2^b \wedge \frac{\omega_\Sigma^{m-1}}{(m-1)!} \ .
\label{2.7}
\end{equation}
But the definition of the complex structure $J_\MM$ and standard manipulations in Hodge theory of K\"ahler manifolds (see for example \cite{Voisin}, page 150) also yield the formula
\begin{equation}
(J_\MM \,\dA ) \wedge \frac{\omega_\Sigma^{m-1}}{(m-1)!} \ = \ \ast_\Sigma\, \dA \ .
\label{2.8}
\end{equation}
This finally implies that 
\begin{align}
\omega_\MM (\dA_1 + \dphi_1 ,\: J_\MM (\dA_2 + \dphi_2) ) \ &= \ \int_\Sigma \omega_X (\dphi_1 , J_X \dphi_2)\, \frac{\omega_\Sigma^m}{m!} \ +\ \frac{1}{4e^2}\, k_{ab}\, \dA_1^a \wedge \ast_\Sigma \dA_2^b  
\nonumber \\
&= \ g_\MM (\dA_1 + \dphi_1 ,\: \dA_2 + \dphi_2) \ ,   \nonumber
\end{align}
and confirms that $\omega_\MM$ is the K\"ahler form associated to the metric $g_\MM$.
\end{prooff}

\section{Simplication for linear targets}

In this section we will see how formula (\ref{2.1}) can be further simplified when the target $X \simeq \CC^n$ is a complex vector space and $G$ acts through a unitary representation $\rho : G \rightarrow U(n)$. With these choices the fibre bundle $P\times_G X$ becomes simply a rank $n$ complex vector bundle $V \rightarrow \Sigma$. Moreover, if we choose the standard K\"ahler form on $\CC^n$
\[
\omega_{\CC^n} \ = \ \frac{i}{2}  \sum_k   \dd w^k \wedge \dd \bar{w}^k \ ,
\label{3.1}
\]
the moment map $\mu : \CC^n \rightarrow \g^\ast$ is given by the formula 
\begin{equation}
\mu (w) \ = \ \frac{1}{2} [ i w^\dagger \, (\dd \rho) (e_a) \, w \, + \, \tau_a  ] \, e^a  \ ,
\end{equation}
where $\dd \rho : \g \rightarrow u(n)$ is the derivative of the representation $\rho$, and $\tau = \tau_a e^a$ is any element of the dual $\g^\ast$ that annihilates the subspace $[\g , \g]$ of $\g$.
A first observation concerns the cohomology class $[\omega_\MM]$ in $H^2 (\MM )$ of the K\"ahler form.
\begin{prop}
Over the smooth region $\MM_{\rm smooth}$ of the moduli space the cohomology class of the K\"ahler form is 
\begin{equation}
[\omega_\MM ] \ = \  - \int_\Sigma  \frac{\tau_a}{2\, m!} \: [F_{\AAA}^a] \wedge [ \omega_\Sigma^m] \ + \ \frac{1}{4 e^2 (m-1)!}\:  [\: k_{ab}  \, F_\AAA^a \wedge F_\AAA^b\: ] \wedge [\omega_\Sigma^{m-1}] \ .
\label{3.2}
\end{equation}
\label{prop3.1}
\end{prop}
\begin{prooff}
We have only to recall that when $X$ is a vector space acted by a linear representation, deformation invariance implies that two closed $G$-equivariant forms are cohomologous in $H_G^\bullet (X)$ if and only if they coincide at the origin of the vector space (see for example \cite{G-G-K}). This shows that in $H^2_G (\CC^n)$ :
\begin{equation}
[\omega_X^G] _G \ = \ [\omega_{\CC^n}  - e^a  \otimes \mu_a (w) ]_G \ = \ [- e^a \otimes \tau_a / 2]_G \ .
\end{equation}
This is a great simplification, for then we have that
\label{3.3}
\begin{equation}
\ev^\ast [\omega_X^G]_G \ = \ \ev^\ast [- e^a \otimes \tau_a /2 ]_G\ = \ - [F_\AAA^a]\:  \tau_a /2 
\end{equation}
in $H^2 (\MM_{\rm smooth}  \times \Sigma)$, as desired.
\end{prooff}

Imposing an additional condition on the representation $\rho$, it is a remarkable fact that formula (\ref{3.2}) holds also for the differential forms themselves. This has already been recognized in the literature in several instances.
\begin{prop}
Assume that the derivative $\dd \rho : \g \rightarrow u(n)$ of the representation $\rho$ is such that
\begin{itemize}
\item[(i)] the matrix $i 1_{n\times n}$ belongs to the image $\dd \rho (\g)$ \ ;
\item[(ii)] the inner products on the Lie algebras are preserved up to rescaling. 
\end{itemize}
Then over the smooth part of the moduli space
\begin{equation}
\omega_\MM  \ = \  - \int_\Sigma  \frac{\tau_a}{2\, m!}\, F_\AAA^a \wedge \omega_\Sigma^m \ + \ \frac{1}{4 e^2 (m-1)!}\, k_{ab}\,  F_\AAA^a \wedge F_\AAA^b \wedge \omega_\Sigma^{m-1} \ .
\label{3.4}
\end{equation}
\label{prop3.2}
\end{prop}
\vspace{-0.5cm}
\begin{rem}
This formula was first obtained by Perutz \cite{Perutz} in the case of $G=U(1)$ acting by multiplication on $X=\CC$ and $m=1$. A similar formula was found by Biswas and Schumacher in \cite{Biswas-Schumacher} for $G=U(n)$ and the moduli space of solutions of the coupled vortex equations. Our proof is an extension of Perutz's. 
\end{rem}

\begin{prooff}
Using theorem \ref{thm2.1} and the formulae for $\omega_{\CC^n}$ and $\mu_{\CC^n}$, we only need to show that if (i) and (ii) are satisfied, then
\begin{equation}
\int_\Sigma \ev^\ast \:  \Big[\: \frac{i}{2} \sum_k (\dd w^k \wedge \dd \bar{w}_k ) \ -\ \frac{i}{2} w^\dagger (\dd \rho ) (e_a) w \otimes e^a\: \Big] \wedge \omega^n_\Sigma  \ = \ 0 \ .
\label{3.5}
\end{equation}
The first step is an algebraic manipulation to prove that
\begin{equation}
\ev^\ast [w^\dagger \dd \rho (\varphi^a e_a) w ] \ =\ \phi^\dagger \, \dd \rho (\varphi^a e_a) \, \phi \ =\ \frac{1}{2\lambda e^2}  \, {\rm Tr} [\dd \rho \circ C^\dagger_\phi C_\phi (\varphi^a e_a) ] \
\label{3.6}
\end{equation}
as functions on $\MM \times \Sigma$ with imaginary values. Here $C_\phi$ is the matter part of the operator defined in (\ref{2.1.0}) associated with infinitesimal gauge transformations, $C^\dagger_\phi$ is the adjoint operator and $\lambda$ is some positive real number. The general expression for the composition 
\[
C_\phi^\dagger C_\phi \ : \ \Omega^0 (\Sigma ; P\times_{Ad} \g)  \longrightarrow \Omega^0 (\Sigma ; P\times_{Ad} \g) 
\label{3.7}
\]
is written down in \cite{Baptista2} and, in the case of complex vector spaces, reduces to
\begin{equation}
C_\phi^\dagger C_\phi (\varphi^a e_a) \ = \ - e^2\,  k^{ab}\, \varphi^c \, \phi^\dagger \: \big[ \: \dd \rho  (e_c) \dd \rho  (e_b) +  \dd \rho  (e_b) \dd \rho  (e_c) \: \big] \: \phi\, e_a \ .
\label{3.8}
\end{equation} 
From this one computes that
\begin{align}
{\rm Tr} \big[ \: \dd \rho \circ C^\dagger_\phi C_\phi (\varphi^a e_a)\: \big] \ &= \  - e^2\, k^{ab}\, \varphi^c \, \phi^\dagger\:  \big\{ \dd \rho  (e_c) , \,  \dd \rho  (e_b) \big\} \: \phi \: {\rm Tr}\big[\dd \rho (e_a)\big]  \label{3.9} \\
&= \ -ie^2 \, \phi^\dagger\: \big\{B, \, \dd \rho (\varphi^c e_c) \big\}\:  \phi \ ,    \nonumber
\end{align}
where $\{ \cdot , \cdot \}$ is the anti-commutator and 
\[
B \ = \  k^{ab} \, (\dd \rho) ( e_b) \, {\rm Tr}\big[-i \dd \rho (e_a)\big]  \ .
\label{3.11}
\]
Now observe that assumption $(ii)$ implies that, for all $e_a$, 
\begin{align}
{\rm Tr}[ B^\dagger \dd \rho (e_c)] \ &= \ k^{ab}\,  {\rm Tr}\big[ \dd \rho  (e_b)^\dagger  \,  \dd \rho  (e_c) \big] \ {\rm Tr}\big[-i \dd \rho (e_a)\big]    \label{3.12}  \\
&= \ k^{ab} \, \lambda\, k_{bc}\, {\rm Tr}\big[-i\, \dd \rho (e_a)\big] \ = \  \lambda\,  {\rm Tr}\big[(i \, 1_{n\times n})^\dagger  \dd \rho (e_c)\big]     \nonumber
\end{align}
for some positive real $\lambda$. This means that $B - i \lambda 1_{n\times n}$ belongs to the subspace of $u(n)$ orthogonal to $\dd \rho (\g)$. But $B$ belongs to $\dd \rho (\g)$ and, by assumption (i), so does $i 1_{n\times n}$, hence the vector $B - i 1_{n\times n}$ necessarily vanishes in $u(n)$. This shows that $B$ is proportional to the identity matrix, and hence (\ref{3.9}) leads to the equality (\ref{3.6}).

The second step in the proof is to consider the gauge part $C_A$ of the infinitesimal gauge transformations (\ref{2.1.0}). Using the formulae given in \cite{Baptista2}, we have that
\begin{align}
{\rm Tr} \big[ \; \dd \rho \circ C_A^\dagger C_A (\varphi^a e_a) \; \big] \  &= \ {\rm Tr} \Big\{ \:  \dd \rho  \big[\; - g_\Sigma^{\mu \nu} \; (\partial_\mu \partial_\nu \varphi^a - \Gamma_{\mu\nu}^\lambda  \partial_\lambda \varphi^a )\; e_a \big]  \: \Big\}   \label{3.14} \\
&= \  \dd^\ast \dd  \Big\{ \: {\rm Tr} [\: \dd \rho (\varphi^a e_a ) \:]  \: \Big\}  \ ,   \nonumber
\end{align}
where $\dd^\ast \dd$ is the Hodge laplacian acting on functions on $\Sigma$. Given that this term has a vanishing integral over $\Sigma$, we conclude that 
\begin{equation}
\int_\Sigma  {\rm Tr} \circ \dd \rho  \circ  C_{(A, \phi)}^\dagger C_{(A, \phi) } (\varphi)  \ \omega_\Sigma^m \ = \  2 \lambda e^2 \int_\Sigma \phi^\dagger\,  \dd \rho (\varphi)\,  \phi  \  \omega_\Sigma^m
\label{3.16}
\end{equation} 
for all sections $\varphi$ in $\Omega^0 (\Sigma ; P\times_{Ad} \g)$. Using then the definition of the pull-back $\ev^\ast$ and the Weil homomorphism in (\ref{2.1.5}),  and using also formula (\ref{2.1.2}) for the curvature $F_\AAA$, it is clear that as a 2-form on $\MM$ acting on a tangent vector $(\dA_1 +\dphi_1 , \, \dA_2 +\dphi_2)$,  
\begin{align}
\int_\Sigma   \ev^\ast \big[ \: w^\dagger \dd \rho ( e_a) w   \otimes e^a \: \big] \wedge \omega_\Sigma^m  \ &=   
\int_\Sigma    \phi^\dagger\, \dd \rho ( e_a)\,  \phi   \ F_\AAA^a \wedge \omega_\Sigma^m   \  
\label{3.18}   \\
&=  \  \frac{1}{2 \lambda e^2}  \int_\Sigma   {\rm Tr} \circ \dd \rho (\psi) \ \omega_\Sigma^m \ , \nonumber
\end{align}
where $\psi$ is the section of the adjoint bundle $P\times_{\rm Ad} \g  \rightarrow \Sigma$ defined in (\ref{2.1.3}). But in the case of the linear target $X = \CC^n$ this definition simplifies, and we get
\begin{align}
{\rm Tr} \circ \dd \rho (\psi)\ &= \  - 2\, e^2\, k^{ab} \; {\rm Tr}\big[\dd \rho (e_a)\big] \;   \big[ \: \dphi_2^\dagger \,  (\dd \rho ) (e_b)   \, \dphi_1  + {\rm c.c.}\: \big]   \label{3.19}  \\
&= \ -2\,i \,e^2  \big(\: \dphi_2^\dagger\, B \, \dphi_1 \;  -\;  \dphi_1^\dagger\, B \, \dphi_2 \: \big) \ = \ 2\, \lambda\, e^2 \, \big(\:  \dphi_2^\dagger\, \dphi_1 \: -\:  \dphi_1^\dagger \,\dphi_2 \: \big)   \label{3.20}    \nonumber \\
&= \ 2\, \lambda\, e^2 \,  \ev^\ast (\sum_k  \dd w^k \wedge \dd \bar{w}^k ) \ (\dA_1 +\dphi_1 , \, \dA_2 +\dphi_2 ) \ . \nonumber
\end{align}  
Finally, bringing together (\ref{3.18}) and (\ref{3.5}), we conclude that formula (\ref{3.4}) in the statement of the proposition is true.
\end{prooff}

 \section{The vortex K\"ahler class in abelian linear models}

The formulae for the K\"ahler form $\omega_\MM$ that we have obtained so far all depend on the curvature $F_\AAA$ of the universal connection, and so are not particularly explicit. But while this lack of explicitness is unfortunate, it is as far as one can expect to go at such a general level, because in most cases we know very little about the vortex moduli space $\MM$, sometimes not even knowing whether it is a non-empty set. There are, however, a few notable exceptions to this generic ignorance. The main example is the vortex moduli space in abelian gauged linear sigma-models, i.e. when we pick the target $X$ to be a complex vector space acted by a torus through a linear representation. This example is important in physics \cite{Witten, Morrison-Plesser} and a rather concrete description of $\MM$ is known \cite{Morrison-Plesser, Werheim}. In this section we will use that description to give an equally concrete  formula for the cohomology class $[\omega_\MM]$ of the K\"ahler form for abelian linear models.

$\ $

To fix notation, we now take $\Sigma$ to be a compact Riemann surface and the target $X$ to be $\CC^n$ acted be a linear representation $\rho$ of the $k$-torus with integral weights, or charges, $Q_j^a$, where the indices run as $1\leq j \leq n$ and $1\leq a \leq k$. We take $P \rightarrow \Sigma$ to be a $T^k$-principal bundle of integral degree $\deg P  \in {\mathbb Z}^k$ and define $L_j =  P\times_{\rho_j} \CC$ to be the line-bundle over $\Sigma$ associated with the restriction of the representation to the $j$-th copy of $\CC$ inside $\CC^n$. Then
\[
\deg (L_j) \ = \ \sum_{a=1}^k  Q_j^a \; (\deg P)_a \ .
\label{4.1}
\]
With these conventions $\phi$ is a section of the vector bundle $\oplus^n_{j=1} L_j$ over $\Sigma$, and the vortex equations read
\begin{align}
&\bar{\partial}^A  \phi \ = \ 0     \label{4.2} \\
&\ast F_A^a \,- \, e^2   \Big[  \big( \: \sum_j   Q^a_j  |\phi^j|\: \big)  - \tau^a \Big] \ = \ 0   \qquad a=1, \ldots , k  \ ,  \nonumber
\end{align}
where the $\tau^a$'s are fixed real constants. The moduli space of solutions to these equations depends on the values of the constants $\deg P$, $Q_j^a$ and $\tau^a$, and as these values change it can range from anything as interesting as the empty set to a complicated singular toric variety. In favourable cases, however, the moduli space $\MM$ can also be a smooth complex manifold, and this is the case we will treat in this section. The conditions on the constants $Q_j^a$ and $\tau^a$ that ensure the smoothness of $\MM$ are similar to the conditions that ensure the smoothness of the symplectic quotient $\mu^{-1} (0) / T^k$. They can be stated as follows (see \cite{Werheim} and the appendix).

\begin{assumption (i)}
The two vectors $\tau$ and $\tau - 2 \pi (\deg P) / (e^2 \, {\rm Vol} \,\Sigma )$ in the Lie algebra ${\mathfrak t}^\ast \simeq \RR^k$ can be written as a linear combinations $\sum_{j\in I} c^j  Q_j$ with positive coefficients for some index set $I \subset \{ 1, \ldots , n \}$ and, furthermore, for all such $I$'s the set of weight vectors $\{ Q_j : j\in I  \}$ generates the integer lattice ${\mathbb Z}^k$ in $\RR^k$.  
\end{assumption (i)}
To this requirement we also add:

\begin{assumption (ii)}
For all $j=1 , \ldots , n$ the integers $d_j = \deg L_j = Q_j^a \,  (\deg P)_a$ are positive and bigger than $2g-2$ -- the Euler characteristic of $\Sigma$.
\end{assumption (ii)}
 The second assumption is here to guarantee that $\MM$, besides being smooth, has the following simple description.

\begin{theorem}[\cite{Morrison-Plesser, Werheim} and appendix]
Under assumptions (i) and (ii) the vortex moduli space $\MM$ is a smooth fibre-bundle over the $k$-fold cartesian product of the jacobian $J_\Sigma$. The fibre of $\MM \rightarrow \times_k J_\Sigma$ is isomorphic to the toric manifold $F_{d_j , \tau}$ of positive dimension $ d = \sum_j (1-g+ d_j ) - k$ defined as $\nu^{-1} (0) /  T^k$, where 
\begin{itemize}
\item[(a)] the torus $T^k$ acts on $\CC^{d+k}$ with the same weight $Q_j^a$ on the coordinates $w_{j, 1} , \ldots , w_{j, 1-g+d_j}$ , for all $j = 1, \ldots , n$;

\item[(b)] $\nu$ is the moment map $\CC^{d+k} \rightarrow \RR^k$ of this action, given by
\begin{equation}
\nu (w) \ = \ -\frac{1}{2}  (\sum_{j=1}^n  \sum_{l_j =1}^{1-g + d_j}   Q_j  \,  | w_{j, l_j} |^2 ) \: - \: \tau  \ .
\label{4.4}
\end{equation}
\end{itemize} 
\label{thm4.1}
\end{theorem}

\begin{rem}
As is well-known, the jacobian manifold of a Riemann surface of genus $g$ is isomorphic to $T^{2g}$. In the special case where $\Sigma$ is the 2-sphere the jacobian is a point, and therefore the moduli space $\MM$ coincides with the toric manifold $F_{d_j , \tau}$.
\end{rem}
\begin{rem}
The description of $\MM$ contained in theorem \ref{thm4.1} is valid in more general circumstances. For example if the weights $Q^a_j$ are such that the quotient $F_{d_j , \tau} = \nu^{-1} (0) /  T^k$ is an orbifold, then although $\MM$ will not be smooth, it will still be a fibration over $\times_k J_\Sigma$ with (singular) fibre $F_{d_j , \tau}$.

\end{rem}

\vspace{.3cm}

We will not describe here how theorem \ref{thm4.1} can be proved. A sketch of the proof is given in the appendix and the original arguments can be found in references \cite{Morrison-Plesser, Werheim}. One important point for us, though, is that the fibre bundle $\MM$ described in the theorem is constructed as the quotient of a certain vector bundle $V \rightarrow \times_k J_\Sigma$ by a $(\CC^\ast)^k$-action on the fibre. In fact, the bundle $V$ splits as the direct sum $V = \oplus_j V_j$ of natural complex vector bundles $V_j \rightarrow \times_k J_\Sigma$ of rank $1-g + d_j$. If we then let $(\CC^\ast)^k$ act on the fibres of each $V_j$ by multiplication with weight $Q_j$, we get a global action on the sum $V$ that on each fibre looks like the action $(a)$ of theorem \ref{thm4.1}. The moduli space $\MM$ is then the quotient of the stable set by this $(\CC^\ast)^k$-action, or equivalently, the quotient of $\nu^{-1} (0)$ by the real $T^k$-action.

This construction is relevant for our purposes because it allows us to define a set of natural cohomology classes $\eta_a$ on the moduli space. In fact, observing that $\nu^{-1} (0) \rightarrow \MM$ is a principal $T^k$-bundle, we can define $\eta_a$ to be the first Chern class of the associated line-bundle $\LL_a = \nu^{-1} (0) \times_{U(1)^a} \CC \rightarrow \MM$. (Here the notation means that $T^k$ acts on $\CC$ by simple multiplication of the $a$-th $U(1)$-factor inside $T^k$.) Thus, for the record, 
\begin{equation}
\eta_a \ := \ c_1 (\LL_a)  \qquad \quad{\rm in} \quad H^2 (\MM ; {\mathbb Z}) \quad {\rm for} \ \ a= 1, \ldots , k .
\label{4.5}
\end{equation}
These cohomology classes are standard in toric geometry and are known to generate the cohomology ring of the toric quotient $F_{d_j , \tau}$. Thus, in the case of genus zero, the $\eta_a$'s generate the whole cohomology of $\MM$. For higher genus they generate the cohomology of the toric fibres of $\MM \rightarrow \times_k J_\Sigma$.

In the higher genus case there is another set of natural cohomology classes on $\MM$. These are the pull-backs of the cohomology classes of the jacobians $J_\Sigma \simeq T^{2g}$ by the projection $\MM \rightarrow \times_k J_\Sigma$. If for each jacobian inside the product $\times_k J_\Sigma$ we call $\theta_a$ the Poincar\'e-dual of the theta divisor, then after pulling-back to $\MM$ we get the natural set of classes $\theta_a \in H^2 (\MM ; {\mathbb Z})$. Using both these sets of cohomology classes we can now state the main result of this section.

\begin{theorem}
For the vortex moduli space $\MM$ described in theorem \ref{thm4.1},  the cohomology class of the vortex K\"ahler form is
\begin{equation}
[\omega_\MM] \ = \ 2\pi \sum_{a=1}^k  \Big(\frac{\tau_a}{2} \, {\rm Vol}\, \Sigma  - \frac{\pi}{e^2} (\deg P)_a \Big) \, \eta_a  \ + \ \frac{\pi}{e^2} \, \theta_a 
\label{4.6}
\end{equation} 
inside $H^2 (\MM ; \RR)$.
\label{thm4.2}
\end{theorem}

\begin{rem}
This result extends to more general abelian gauged linear sigma-models the formula found in \cite{Manton2, Perutz} for the classical abelian Higgs model. Our formula is formally very similar to the latter, the main difference being that for non-trivial charges $Q_j^a$ the interpretation of the classes $\eta_a$ is more evolved. Note also that in \cite{Manton2} the K\"ahler form $\omega_\MM$ is normalized so that the global $2\pi$ factor disappears.
\end{rem}

\vspace{.2cm}

\begin{prooff}
To prove this result we use proposition \ref{prop3.1} and examine carefully the class $[F_\AAA^a]$ defined by the curvature of the connection $\AAA$ on the universal bundle $\PP \rightarrow \MM \times \Sigma$.
The first thing to observe is that the group $\GG$ of gauge transformations acts freely on the full set of vortex solutions. This is so because only constant gauge transformations preserve the connection $A$, and, due to the holomorphy condition $\bar{\partial ^A \phi} =0$ and assumption $(i)$, the $T^k$-stabilizer of $\phi(z)$ is trivial for a generic point $z \in \Sigma$. This observation actually belongs to the proof of theorem \ref{thm4.1}, so we need not justify it here further; in any case, it shows that both $\MM = \VVV / \GG $ and $\PP = (\VVV \times P) / \GG$ are globally smooth. Now, as described in section 2.1, the construction of the universal bundle $\PP$ implies that its restriction to $[A, \phi] \times \Sigma$ is isomorphic to $P \rightarrow \Sigma$, and so
\begin{equation}
\frac{-1}{2\pi} \, [F_\AAA^a]\:  | _{[A , \phi] \times \Sigma} \ = \ \frac{-1}{2\pi} \, [F_A^a] \ = \ (\deg P)_a  \qquad {\rm in} \ {\mathbb Z} \simeq H^2 (\Sigma ; {\mathbb Z}) \ .
\label{4.7}
\end{equation}    
On the other hand, if we pick any point $p \in \Sigma$ and define the subgroup of gauge transformations
\[
\GG_p \ := \ \{ g:\Sigma \rightarrow T^k \ {\rm such \ that \ } g(p)=(1, \ldots , 1) \ \} \ ,
\label{4.8}
\]
it is clear that $\GG$ splits as $\GG_p \times T^k$ and that 
\begin{align}
\PP \: |_{\MM \times \{ p\}} \ &= \ (\VVV \times  P_p )\:  /\:  \GG \ = \ [(\VVV \times P_p) / \GG_p ] \: / \: T^k  
\label{4.9} \\
&= \ [(\VVV / \GG_p ) \times P_p] \: / \: T^k \ \simeq \ \VVV \: / \: \GG_p \ ,  \nonumber
\end{align}
where in the last step we have used that the $T^k$-action on the fibre $P_p$ is free and transitive. So we conclude that the restriction of $\PP$ to $\MM \times \{ p\}$ is isomorphic to the $T^k$-bundle $\VVV / \GG_p \rightarrow \MM$.

Consider now the space of holomorphic sections
\[
\BB \ = \ \{  (A, \phi ) : \   \bar{\partial}^A \phi =0  \}
\label{4.11}
\]
and the open subset $\BB^\ast \subset \BB$ where the group $\GG^\CC$ of complex gauge transformations acts freely. As sketched in the appendix, the proof of theorem \ref{thm4.1} defines the total space of the vector bundle $V \rightarrow \times_k J_\Sigma$ as the quotient of $\BB$ by the subgroup $\GG^\CC / (T^k)^\CC \simeq \GG_p^\CC$. The natural $(\CC^\ast)^k = (T^k)^\CC$ action on $V$ then comes from the residual $(\CC^\ast)^k$ action on $\BB / \GG_p^\CC$, and this is free exactly on $\BB^\ast  / \GG_p^\CC$. But the same proof also says that any $(A, \phi)$ in $\BB^\ast$ is in the $\GG^\CC$-orbit of a vortex solution in $\VVV$, and that this solution is unique up to real gauge transformations. This means not only that there is an identification of the full quotients $\MM = \VVV / \GG \simeq  \BB^\ast / \GG^\CC$, but also that $\VVV / \GG_p$ can be identified with any submanifold of $\BB^\ast / \GG_p^\CC$ of real codimension $k$ that intersects once all the $(\CC^\ast)^k$-orbits in $\BB^\ast / \GG_p^\CC$.
Finally,  through the identification of $\BB^\ast / \GG_p^\ast$ with the stable subset of $V$ mentioned before, we see that one such submanifold is exactly $\nu^{-1} (0) \subset V$, where $\nu :V \rightarrow \RR^k$ is the moment map defined fibrewise by (\ref{4.4}). Hence the $T^k$-bundle $\VVV / \GG_p \rightarrow \MM$ is isomorphic both to $\PP\ |_{\MM\times {p}}$ and to $\nu^{-1} (0) \rightarrow \MM$. The conclusion is then that 
\begin{equation}
\frac{-1}{2\pi} \, [F_\AAA^a]\:  | _{\MM \times \{ p\}} \ = \ c_1 ( \PP\: |_{\MM\times {p}}  \times_{U(1)^a} \CC ) \ = \ c_1(\LL_a) \ = \ \eta_a  
\label{4.12}
\end{equation}
in $H^2 (\MM ; {\mathbb Z})$.

Having computed the restriction of the curvature class $[F_\AAA]$ to both $[A, \phi] \times \Sigma$ and  $\MM \times \{p\}$, the only remaining task is to describe the components of $[F_\AAA]$ that have one leg in $\Sigma$ and one leg in $\MM$. For this we consider the standard set $\{ \alpha_l , \alpha_{l+g}  : \: l=1, \dots , g \}$ of generators  of $H^1 (\Sigma , {\mathbb Z})$ and write
\begin{equation}
\frac{-1}{2\pi} \: [F_\AAA]_{\rm mixed} \ = \ \sum_{j=1}^g \ \gamma^a_j \wedge \alpha_j \ + \  \gamma_{j+g}^a \wedge \alpha_{j+g}
\label{4.13}
\end{equation}
in the cohomology $H^2 (\MM \times \Sigma ; {\mathbb Z})$. Then the $\gamma$'s are uniquely determined elements of $H^1 (\MM ; {\mathbb Z})$ that contain all the relevant information about $[F_\AAA]_{\rm mixed}$. Leaving them unspecified for the moment, we can press on with the K\"ahler class computation and substitute in the integral of proposition \ref{prop3.1} (with $m=1$) the various components (\ref{4.7}),  (\ref{4.12}), (\ref{4.13}) of the curvature. This leads to the expression 
\begin{equation}
[\omega_\MM] \ = \  \
\pi \; \sum_a \; \Big\{ \: \big[    \tau^a \, ({\rm Vol} \Sigma ) \ -\ \frac{2\pi}{e^2}\, (\deg P)^a \big]\; \eta_a \ + \ \frac{2\pi^2}{e^2} \; \sum_{j=1}^{g} \; \gamma^a_j \wedge \gamma^a_{j+g}  \; \Big\} \ ,
\end{equation}
and is almost the formula stated in the theorem. The only task left is to identify the class $\sum_j  \gamma^a_j \wedge \gamma^a_{j+g}$ with the class $\theta_a$ in $H^2 (\MM ; {\mathbb Z})$.  This can be done through the following indirect observation.

Looking at the formulae for the action of the curvature form $F_\AAA$ on tangent vectors, one sees that the second line in (\ref{2.1.2}) depends only on $v \in T\Sigma$ and on $\dA$, not on $\dphi$. This means that the leg of $(F_\AAA)_{\rm mixed}$ resting on $\MM$ actually rests only on the base of the fibration $\MM \rightarrow \times_k J_\Sigma$, because it is the base that parametrizes the different complex structures on the $L_j$'s (which, recall, determine and are determined by the different connections $A$). In other words, since the mixed part of $F_\AAA$ does not depend on $\dphi$, it is the same as in the vortex equations with target $X={\rm point}$, and this is just the pure gauge theory with moduli space $\MM_{\rm pure \ gauge} = \times_k J_\Sigma$. 
This means that the $\gamma^a$'s in formula (\ref{4.13}) are actually classes in $\times_k J_\Sigma$ pulled back to $\MM$ and that, as classes in $\times_k J_\Sigma$, they do not really depend on the matter part of the gauged linear model we are working on. Comparing with the simpler abelian Higgs model of \cite{Manton2, Perutz, Bertram-al}, we conclude that the class $\sum_l \gamma_l^a \wedge \gamma_{l+g}^a$, as a class in the $a$-th jacobian $J_\Sigma \simeq T^{2g}$, is just the standard integral symplectic form on this torus, or in other words it is the Poincar\'e-dual $\theta_a$ of the theta-divisor in $J_\Sigma$. This concludes the proof.
\end{prooff}

\section{Application: volume of simple moduli spaces}

\subsection{Volume of spaces of abelian semi-local vortices}

On any given compact K\"ahler manifold there are several interesting metric quantities that can be computed with a very limited knowledge of the local details of the metric. Two important examples are the overall volume of the manifold and the global integral of the scalar curvature, which depend solely on the K\"ahler class of metric and on the cohomology ring of manifold. Here we are interested in exploiting these facts in the case of vortex moduli spaces and metrics. Observe in particular that for the abelian models studied in the last section our odds look quite promising, for we already have a  topological description of the moduli space (theorem \ref{thm4.1}) and of the K\"ahler class of the metric (theorem \ref{thm4.2}). 
It turns out, however, that for the general toric fibrations $F_{d_j, \tau} \rightarrow \MM \rightarrow \times_k J_\Sigma$ of section 4, the intersection combinatorics on $\MM$ can be complicated, and so the volume computation is not straightforward.
 A nice and easy exception, nevertheless, occurs when we consider the simple case of the vortex moduli space for group  $G= U(1)$ acting by scalar multiplication on the target $X=\CC^n$. In this case all the toric manifolds $F_{d_j , \tau}$  become projective spaces, which have a simple cohomology ring, and so we can easily carry out the computations to the end. 

  The volume computations presented here extend several results that can be found in the literature. These include the case $n=1$ and arbitrary $\Sigma$ calculated in \cite{Manton2}, and the case of arbitrary $n\geq 1$ and special $\Sigma = T^2$ calculated in \cite{Eto-al2} through a heuristic brane construction. A third method to compute the volume of $\MM$ in the case $\Sigma = \CC {\mathbb P}^1$ and $n=1$ was described in \cite{Romao}. The scalar curvature integrals presented here, on the other hand, had not been considered before.

$\ $

As described above, in this section we will take $\Sigma$ to be a compact Riemann surface and consider the special case of vortices with group $U(1)$ acting by scalar multiplication on the target $\CC^n$. These are sometimes called abelian semi-local vortices. In the notation of section 4, they correspond to the choice $k=1$ and charges $Q_1^1 = \cdots = Q_n^1 =1$. 
Applied to this particular case, theorem \ref{thm4.1} and its proof say that if $\deg P = d$ is positive and satisfies $d > 2g-2$, then for sufficiently big volume of the surface $\Sigma$ the moduli space $\MM$ of vortex solutions is isomorphic to a projective bundle. More specifically, there exists a complex vector bundle $V \rightarrow J_\Sigma$ of rank $n(1-g + d)$ over the jacobian of $\Sigma$ such that
\begin{equation}
\MM \ \simeq \ {\mathbb P} (V) \ \rightarrow \ J_\Sigma \ .
\label{5.1}
\end{equation}
The total Chern class of this vector bundle has the very simple form $c(V) = e^{- n \theta}$, where $\theta$ has the same meaning as in section 4, i.e. it is the degree 2 cohomology class on $J_\Sigma$ that is Poincar\'e-dual to the theta divisor \cite{Bertram-al}. Also, with these choices, the formula in theorem \ref{thm4.2} for the cohomology class of the vortex K\"ahler form specializes to 
\begin{equation}
[\omega_\MM] \ = \  \pi \Big( \tau \, {\rm Vol}\, \Sigma  - \frac{2 \pi}{e^2}\, d \Big) \, \eta  \ + \ \frac{2 \pi^2}{e^2} \, \theta \ .
\label{5.2}
\end{equation}
Observe that this description of the moduli space is valid only if the constant $\tau e^2\, {\rm Vol}\, \Sigma$ is bigger than $2\pi d$ (assumption (i) of section 4). If it is smaller, it is well-known that $\MM$ is actually the empty set, as can be easily recognized by integrating over $\Sigma$ the second vortex equation. Now, to make further computations with this expression for $[\omega_\MM]$, it is important to understand clearly the meaning of the class $\eta \in H^2 (\MM , {\mathbb Z})$. Firstly, define $V^\ast := V\setminus \{{\rm zero\ section}\}$ and consider it as a principal $\CC^\ast$-bundle over the jacobian. A careful inspection of the definition of $\eta$ in section 4 and of the proof of theorem \ref{thm4.2}, shows that $\eta$ is the first Chern class of the associated bundle $V^\ast \times_{\CC^\ast} \CC$. But this bundle is defined by the simple equivalence relation $(v, w) \sim (\lambda v , \lambda w)$ in the product $V^\ast \times \CC$, and it is a standard fact that this relation defines nothing more than the dual of the tautological line-bundle $L \rightarrow {\mathbb P} (V)$ over the projectivization. Thus, going back to definition (\ref{4.5}), we conclude that ${\mathcal L} \simeq L^\ast$ and that the class $\eta$ is just the first Chern class of the anti-tautological bundle $L^\ast \rightarrow {\mathbb P}(V)$. 

As a last preparatory point we also need to write down the first Chern class of the moduli space $\MM$. This will be necessary for the evaluation of the scalar curvature integral. General results for projective bundles ${\mathbb P}(V) \rightarrow J$ say that  the total Chern classes satisfy 
\[
c[T {\mathbb P}(V)] \ = \ c (V \otimes L^\ast) \ c(TJ) \ .
\]
Using the fact that in our case the base of the fibration is the jacobian -- a group manifold with trivial tangent bundle -- this formula then implies that
\begin{align}
c_1 ({T \MM}) \ &= \ ({\rm rank}\, V) \ c_1 (L^\ast) \ + \ c_1 (V)   \label{5.3}   \\
&= \ n(d+1-g) \: \eta \ - \ n\: \theta .  \nonumber
\end{align}

Having registred these general formulae, we can now turn our attention to the integrals that directly concern us, namely
\begin{equation}
{\rm Vol}\,  \MM \ = \  \int_\MM  {\rm vol}_\MM \ = \ \frac{1}{r!} \int_\MM  [\omega_\MM]^r   
\label{5.4}
\end{equation}
for the volume, and
\begin{equation}
 \int_\MM  s \ {\rm vol}_\MM \ = \ \frac{2\pi}{(r-1)!} \int_\MM c_1(\MM) \wedge [\omega_\MM]^{r-1} \ 
\label{5.5}
\end{equation}
for the total scalar curvature. Here $r = g + n(d+1 - g) -1$ is the complex dimension of the moduli space and, in both expressions, the last equalities are well-known facts of K\"ahler geometry. 
Now, calling $\pi$ the natural projection on the fibre bundle $\MM \rightarrow J_\Sigma$, standard integration over the fibre yields the first equality of 
\begin{equation}
\int_\MM  \eta^{r-i} \wedge \pi^\ast \theta^i \ = \ \int_{J_\Sigma} (\pi_\ast \eta^{r-i}) \wedge \theta^i \ =  \ \frac{n^{g-i} g!}{ (g-i)!}\ .
\label{5.6}
\end{equation}
The last equality, on its turn, is a consequence of the following two facts. Firstly, using our previous interpretation of the class $\eta$, a simple manipulation explained in \cite{Bertram-al} says that
\begin{equation}
\pi_\ast  (\eta^{r-i}) \ = \ (n \, \theta)^{g-i} / (g-i)! \ .
\label{5.7}
\end{equation} 
Secondly, the identification of $\theta = \sum_l  \gamma_l \wedge \gamma_{l+g}$ with the standard symplectic form on the jacobian torus (explained at the end of section 4) leads straightforwardly to
\[
\int_{J_\Sigma} \theta^g \ = \ g! \ .
\]
And this is all we need to know. Combining (\ref{5.6}) with expressions (\ref{5.2}) and (\ref{5.3}) for the K\"ahler and first Chern classes of $\MM$, a final direct computation yields
\begin{theorem}
Consider the vortex equations with group $U(1)$ acting on the target $\CC^n$ by scalar multiplication, and assume that the degree $d = c_1 (P)$ is positive and satisfies $\tau e^2 \,  ({\rm Vol}\, \Sigma)  / (2\pi)  > d > 2g-2$. Then the vortex metric $g_\MM$ on the moduli space has total volume
\begin{equation}
{\rm Vol}\,  \MM \ = \    \pi^r \ \sum_{i=0}^g \  \frac{g!\:  n^{g-i}}{i! \: (r-i)! \: (g-i)!} \ \Big(\frac{2\pi}{e^2}\Big)^i \ \Big( \tau \,  {\rm Vol}\, \Sigma  -  \frac{2 \pi}{e^2}\,  d \Big)^{r-i}
\label{5.8}
\end{equation}
and total scalar curvature (or Einstein-Hilbert action)
\begin{equation}
 \int_\MM  s \ {\rm vol}_\MM \ = \ (2\pi)^r \ \sum_{i=0}^{g} \  \frac{g!\:  n^{g-i} \: \big(  r+1 - 2g +i \big)}{i! \: (r -1-i)! \: (g -i)!}  \ \Big(\frac{\pi}{e^2}\Big)^i \ \Big(   \frac{\tau}{2}  {\rm Vol}\, \Sigma  -  \frac{\pi}{e^2} d \Big)^{r-1-i} ,
\label{5.9}
\end{equation}
where the integer $r$ is defined by $r = g + n(d+1 - g) -1$. If the degree $d$ is bigger than $\tau e^2 \,  ({\rm Vol}\, \Sigma)  / (2\pi)$, then $\MM$ is the empty set.
\label{thm5.1}
\end{theorem}
\begin{rem}
To compare our volume result with the ones obtained in  \cite{Manton2} and \cite{Eto-al2}, we first observe that in both these references the natural vortex metric (\ref{1.1}) is defined with a prefactor of $1/ \pi$, chosen to make the global factor $\pi^r$ disappear in the end result. Then with the choices $\tau =1$ and $e^2 = 1/2$, our formula (\ref{5.8}) reduces exactly to the result presented in \cite{Manton2} for $n=1$. Moreover, in the genus one case, (\ref{5.8}) also reduces to the formula of \cite{Eto-al2}, showing that for $\Sigma = T^2$ the heuristic brane methods remarkably predict the correct result.
\end{rem}
\begin{rem}
Having computed both the total volume and the total scalar curvature of the vortex metrics $g_\MM$, it is possible to obtain the explicit values of the Yamabe functional $I(g_\MM) =   (\int_\MM  s \ {\rm vol}_\MM) \: / \: ({\rm Vol}\,  \MM)^{r / (r-1)}$ for these metrics. For example in the simplest case of genus zero the moduli space is $\MM = \CC {\mathbb P}^r$ and 
 we have $I(g_\MM)= 2 \pi r (r+1) \, / \,  (r!)^{1/ r}$, which no longer depends on the volume of $\Sigma$ and on the constants $\tau$ and $e^2$.
\end{rem}

\subsection{Volume of spaces of holomorphic curves}

The vortex moduli spaces that we have been considering in this section are very closely related to the spaces of holomorphic maps from the Riemann surface to projective spaces. In fact, as explained for example in \cite{Bertram-al}, for each degree $d \geq 2g$ the moduli space $\MM_{\Sigma, d}$ of vortices with group $U(1)$ and target $\CC^n$ is a smooth compactification of the space $\HH_{\Sigma, d}$ of holomorphic maps $\Sigma \rightarrow {\mathbb CP}^{n-1}$ of the same degree. Moreover, each vortex solution $(A, \phi)$ such that $\phi(z)$ never vanishes as a section of $\oplus^n L \rightarrow \Sigma$ defines in a natural way a holomorphic map $\bar{\phi}: \Sigma \rightarrow {\mathbb CP}^{n-1}$ by the formula
\begin{equation}
\bar{\phi} (z) \ = \ \big[ \phi_1 (z), \ldots , \phi_n (z) \big]  \ ,
\label{5.2.1}
\end{equation}
where the components $\phi_j$ are written using any local holomorphic trivialization of $L$ (holomorphic with respect to the complex structure on $L$ induced by $A$), and using homogeneous coordinates on the projective space. Observe that the map (\ref{5.2.1}) is invariant under complex gauge transformations of the section $\phi$. Calling $\MM^o \subset \MM$ the open dense subset of vortex solutions such that the components of $\phi(z)$ never vanish simultaneously, the formula above then defines a biholomorphism $\MM^o_{\Sigma, d} \simeq \HH_{\Sigma, d}$ (see \cite{Bertram-al} for more details).

$\ $

Once we have these identifications in mind, it is obvious to ask whether the natural $L^2$-metric on the vortex moduli spaces bears any relation to the natural $L^2$-metric on the spaces of holomorphic maps. Looking at the definition (\ref{1.1}) of the vortex metric, it is clear that the hope lies in the limit $e^2 \rightarrow \infty$, for in this case the terms that depend on $\dot{A}$ drop out and leave the bilinear form on the space of tangent vectors determined by
\begin{equation}
\big [ (\dot{A}, \dot{\phi}) , (\dot{A}, \dot{\phi}) \big] \ \longmapsto \ \int_\Sigma |\dot{\phi}|^2 \ {\rm vol_\Sigma} \ .
\label{5.2.2}
\end{equation}
Now, although this quadratic form clearly does not define a metric on the space of all tangent vectors $(\dot{A}, \dot{\phi})$, it is easy to check that if the $(\dot{A}, \dot{\phi})$'s satisfy the linearized version of $\db^A \phi = 0$ and are tangent at a point $(A, \phi)$ where $\phi (z)$ never vanishes, then (\ref{5.2.2}) is actually positive definite. This means that (\ref{5.2.2}) does define a metric on the space of vortex solutions above $\MM^o$, and since this metric is invariant by real gauge transformations, also a metric on $\MM^o$ itself. This metric corresponds to keeping just the second term on the right-hand side of (\ref{1.1}), and as the constant $e^2$ becomes larger it differs less and less from the vortex metric. 

And what is the limit of both these metrics when $e^2$ goes to infinity? To start answering this question, observe that although the vortex solutions change as $e^2$ increases, they only change by a complex gauge transformation. This is so because the usual Hitchin-Kobayashi correspondence tells us that any vortex solution, independently of the value of $e^2$, is complex gauge equivalent to a solution of $\db^A\phi = 0$. In particular this implies that the induced holomorphic map $\bar{\phi}: \Sigma \rightarrow {\mathbb CP}^{n-1}$ of formula (\ref{5.2.1}) does not depend on $e^2$. Consider now the formal limit of the vortex equations (\ref{4.2})as $e^2 \rightarrow \infty$. It reads
\begin{align}
&\bar{\partial}^A  \phi \ = \ 0     \label{5.2.3} \\
&\mu \circ \phi \ = \frac{1}{2}\: \Big(\: \tau - \sum_j   |\phi^j|^2 \: \Big) \ = \ 0  \  ,  \nonumber
\end{align}
and  corresponds to holomorphic sections $\phi$  whose image is entirely contained in the subset $\mu^{-1} (0) \subset \CC^n$. Such a section also descends to a map $\bar{\phi}$, and it is shown in \cite{C-G-S} that for $n > 1$ this correspondence establishes a bijection $\MM^o_\infty \simeq \HH$ between the space of equivalence classes of solutions of (\ref{5.2.3}) and the space of holomorphic curves to ${\mathbb CP}^{n-1}$. Now, if the image $\phi (z)$ is always contained in the zero level $\mu^{-1} (0)$, then it follows from the very definition of the induced metric $\omega_{{\mathbb CP}^{n-1}}$ on the symplectic quotient $\mu^{-1}(0)\, /\, U(1) \simeq {\mathbb CP}^{n-1}$ that the metric on the moduli space $\MM_\infty^o$ determined by (\ref{5.2.2}) corresponds under the identification $\MM^o_\infty \simeq \HH$ to the usual metric on the space of holomorphic maps $\bar{\phi}$ to the K\"ahler target $({\mathbb CP}^{n-1}, \omega_{{\mathbb CP}^{n-1}})$. Thus the hope is that if as $e^2$ increases the vortex solutions on $\MM^o$ would ``nicely'' approach the solutions of (\ref{5.2.3}), then the metric on $\MM^o$ determined by (\ref{5.2.2}) -- and hence also the vortex metric -- would ``nicely'' approach the metric on $\MM_\infty^o$ determined by the same formula, which, as we have seen, under the identifications $\MM^o_\infty \simeq \HH \simeq \MM^o$ is nothing but the natural $L^2$-metric on $\HH$. 
\begin{conjecture}
With the identification $\MM^o \simeq \HH$ described by (\ref{5.2.1}), the  pointwise limit of the vortex metric as $e^2 \rightarrow \infty$ is the natural $L^2$-metric on the space of holomorphic maps $\Sigma \rightarrow {\mathbb CP}^{n-1}$, where the target $\mu^{-1}(0) / U(1) \simeq {\mathbb CP}^{n-1}$ should be endowed with the quotient K\"ahler metric.
\end{conjecture}
It is straightforward to check that for the moment map specified in (\ref{5.2.3}) this quotient metric is just
\begin{equation}
\omega_{{\mathbb CP}^{n-1}} \ = \ \pi  \tau \  \omega_{\rm norm.\: FS} \ ,
\label{5.2.4}
\end{equation}
where $\omega_{\rm norm.\: FS}$ denotes the normalized Fubini-Study form on the projective space (the generator of the integer cohomology).
This conjecture can perhaps be made rigorous through a careful study of the analytic results in \cite{Gaio-Salamon} that describe how the vortex solutions approach the solutions of (\ref{5.2.3}) in the adiabatic limit $e^2 \rightarrow \infty$. Similar results may also be true for vortex equations with other targets and gauge groups.

Pressing forward with this hypothesis, one can then use theorem \ref{thm5.1} and the informal manipulation
\begin{equation}
{\rm Vol}\, \HH \ = \ \int_{\HH} {\rm vol}_\HH \ = \ \int_{\MM^o}\ \lim_{e \rightarrow \infty} ({\rm vol}_\MM) \ = \ \lim_{e \rightarrow \infty} \ ({\rm Vol}\ \MM)
\label{5.2.5}
\end{equation} 
to obtain conjectural formulae for the integrals over the non-compact spaces $\HH$.
\begin{conjecture}
For $n>1$ and $d\geq 2g$, the total volume and the total scalar curvature of the natural $L^2$-metric on the space $\HH$ of degree $d$ holomorphic maps $\Sigma \rightarrow ({\mathbb CP}^{n-1} ,\:\omega_{{\mathbb CP}^{n-1}})$ are given by 
\begin{equation}
{\rm Vol}\  \HH \ =    \  \frac{  n^{g}}{\big[n(d+1-g) +g -1\big]! }  \ \big( \pi  \tau \,  {\rm Vol}\, \Sigma  \big)^{n(d+1 -g) +g -1} \\
\label{5.2.6}
\end{equation}
\begin{equation}
 \int_\HH  s \ {\rm vol}_\HH \ = \  \frac{2\pi \, n^{g} \: \big[  n(d+1 -g ) - g \big]}{\big[ n(d+1-g) +g -2\big]! }  \ \ \big( \pi  \tau \, {\rm Vol}\, \Sigma \big)^{n(d+1-g) +g -2} \ .
\label{5.2.7}
\end{equation}
\end{conjecture}
\begin{rem}
As far as the author is aware, the only instance where these integrals have been computed so far is the direct volume integration in \cite{Baptista0} for the very special case of degree one ${\mathbb CP}^1$-lumps on the sphere. This corresponds to the choices $g = 0$, $d = 1$ and $n = 2$ in the formulae above. Taking into account that in \cite{Baptista0} the Fubiny-Study metrics on both ${\mathbb CP}^1$'s were chosen with volume $\pi$ -- and hence for the target space this corresponds to the choice $\tau = 1$ -- we see that (\ref{5.2.6}) reduces exactly to the value $\pi^6 /6$ obtained there.\footnote{Since the first version of the present paper appeared, Speight has rigorously verified the volume formula (\ref{5.2.6}) also in the case $g=0$, $d=1$ and arbitrary $n$ \cite{Speight}.}
\end{rem}

\section{Samols' localization for vortex metrics}

The definition (\ref{1.1}) of the vortex K\"ahler metric involves an integration over the manifold $\Sigma$, hence its common designation as the $L^2$-metric. In very special cases, however, it turns out that this integral can be localized, and appears as a sum of contributions from complex codimension-1 submanifolds of $\Sigma$. This phenomenon was first studied in detail by Samols in the case of the classical abelian Higgs model over a Riemann surface \cite{Samols}, where the group $U(1)$ acts on the target $\CC$ and the submanifolds of $\Sigma$ are just points. It is not, however, a peculiarity of that single example. In fact, as we will see, one should expect something similar to happen roughly whenever the K\"ahler quotient $X/\!/G$ is a point, or in other words, whenever the complexified $G^\CC$-action is free and transitive on an open dense subset $X^o \subset X$. Here the two main examples that we will keep in mind are the vortex equations with group $G = U(N)$ acting on the space of $N\times N$ square matrices, studied for example in \cite{Bertram-al, A-B-E-K-Y, Hanany-Tong, Eto-al1, Baptista3}, and the abelian equations with $G$ a torus and $X$ a toric manifold \cite{Baptista1}.   

$\ $

This section can be read independently of the previous four, and the notation will be as follows. The local complex coordinates on the base $\Sigma$ and on the target $X$ are, respectively, $\{ z^\alpha\}$ and $\{ w^j\}$. With respect to these the K\"ahler forms are written as $\omega_\Sigma = (i/2) h_{\alpha \bar{\beta}}\, \dd z^\alpha \wedge \dd z^{\bar{\beta}}$ and $\omega_X = (i/2) h_{j \bar{k}}\, \dd w^j \wedge \dd w^{\bar{k}}$. The group $G$ has Lie algebra $\g$ with a basis $\{ e_a \}$, and for each vector $v \in \g$ the corresponding vector field on $X$ induced by the left action is called $\hat{v}$. Observe then that over the open subset set $X^o \subset X$ where $G^\CC$ acts freely and transitively, the complex span of the vectors $\hat{e}_a$ is the full tangent space.

The definition (\ref{1.1}) of the vortex $L^2$-metric $g_\MM$ says that for a tangent vector $(\dA , \dphi)$ orthogonal to gauge transformations (also called a horizontal tangent vector) the $L^2$-norm is given by
\begin{equation}
|\!|\,(\dA , \dphi)\, |\!|_{L^2}^2 \ =\ \int_\Sigma {\mathcal E}(\dA , \dphi)\ {\rm vol}_\Sigma \ = \ \int_\Sigma \ \frac{1}{4e^2} \: k_{ab} \: \dot{A}^a \wedge  \ast \: \dot{A}^b  \ + \  g_X ( \dot{\phi} , \dot{\phi} )\ {\rm vol}_\Sigma  \ .
\label{6.1}
\end{equation}
The localization of this integral follows from the fact that the energy density ${\mathcal E}(\dA , \dphi)\ {\rm vol}_\Sigma$ is actually an exact differential form on the inverse image $\phi^{-1}(X^o) \subset \Sigma$.
\begin{prop}
Let $(A, \phi)$ be a vortex solution such that $\Sigma^o := \phi^{-1} (X^o)$ is an open dense subset of $\Sigma$. Then for any horizontal tangent vector at $(A, \phi)$ to the space of vortex solutions, we have that 
\begin{equation}
{\mathcal E}(\dA , \dphi) \ = \ -\, \frac{2}{e^2} \; h^{\alpha \bar{\beta}} \; \partial_\alpha \: (k_{ab}\, \bar{\lambda}^a\, \dA^b_{\bar{\beta}}) 
\label{6.2}
\end{equation}
over $\Sigma^o$, where the complex $\lambda^a$'s are defined by $\dphi = \lambda^a\, \hat{e}_a$ in $TX$. In particular the density ${\mathcal E}(\dA , \dphi)\; {\rm vol}_\Sigma$ is an exact differential form over $\Sigma^o$.
\label{prop6.1}
\end{prop}
\begin{prooff}
In local coordinates the first vortex equation in (\ref{1.2}) can be written as 
 \begin{equation}
 \partial_\balpha \phi^j \ + A_\balpha^a \; \he^j_a \ = \ 0 \ ,
 \end{equation}
 where the quantities defined on the target $X$ are implicitly being evaluated at the point $\phi (z)$. Its linearization therefore reads
\begin{equation}
\partial_\balpha \dphi^j + A^a_\balpha \, (\partial_k \he^j) \, \dphi^k + \dA^a_\balpha \, \he^j_a \ = \ 0 \ .
\label{6.3}
\end{equation}
Substituting into this expression the definition $\dphi^j = \lambda^a\, \he_a^j$ and using the standard fact that
\[
[\he_a , \he_b] \ = \ -f_{ab}^c \, \he_c \ ,
\]
where the $f_{ab}^c$'s are the structure constants of the Lie algebra, this linearized equation reduces to
\begin{equation}
(\partial_\balpha \lambda^a + \dA_\balpha^a + f_{ab}^c \, A_\balpha^b \, \lambda^c )\; \he^a \ = \ 0 \ ,
\label{6.4}
\end{equation}
which over $\Sigma^o$ implies that
\begin{equation}
\dA_\balpha^a \ = \ - \nabla_\balpha^A \lambda^a \ .
\label{6.5}
\end{equation}
The linearization of the second vortex equation, on its turn, can be combined with the horizontallity condition into the single complex equation \cite{Baptista2}
\begin{equation}
2 \, h^{\alpha \bbeta} \, (\nabla_\alpha^A \dA_\bbeta) \ + \ e^2\,  k^{ab} \, h_{j \bk} \,\,  \overline{\he^k_b}\, \, \dphi^j \ = \ 0 \ .
\label{6.6}
\end{equation}
(Recall that the horizontallity condition, which is the real part of (\ref{6.6}), is called in physics Gauss' law, and just demands orthogonality between $(\dA , \dphi)$ and all vectors tangent to real gauge transformations.)
Using (\ref{6.5}) and (\ref{6.6}) it is then straightforward to check that 
\begin{align}
h^{\alpha \bbeta}\, k_{ab}\, \dA_\alpha^a\, \dA_\bbeta^b \ &= \ - h^{\alpha \bbeta} \, k_{ab}\, \big[ \: \partial_\alpha\, (\bar{\lambda}^a\, \dA^b_\bbeta) \ - \ \bar{\lambda}^a\: (\nabla_\alpha^A \, \dA^b_\bbeta)\: \big]  \nonumber \\
&= \ - h^{\alpha \bbeta} \, \partial_\alpha \, ( k_{ab}\, \bar{\lambda}^a\, \dA^b_\bbeta ) \ - \ \frac{e^2}{2}\: h_{j\bk} \, \overline{\dphi^k} \, \dphi^j \ ,
\label{6.7}
\end{align}
and (\ref{6.2}) readily follows. Finally, defining the real 1-form on $\Sigma^o$
\begin{equation}
\eta \ = \ \frac{2}{e^2} \; k_{ab} \,\bar{\lambda}^a\, \dA^b_\bbeta \; \dd \bar{z}^\beta \ + \ {\rm c.c.} \ ,
\label{6.8}
\end{equation}
standard manipulations in K\"ahler geometry show that
\begin{align}
{\mathcal E}(\dA , \dphi)\; {\rm vol}_\Sigma \ &= \ - \frac{1}{2} \; g_\Sigma (\omega_\Sigma , \dd \eta) \; {\rm vol}_\Sigma \ = \ - \frac{1}{2} \dd \eta \wedge \ast \omega_\Sigma  \label{6.9}  \\
&= \ \dd (- \frac{1}{2}  \eta \wedge \ast \omega_\Sigma ) \ , \nonumber
\end{align}
where we have used that $\omega_\Sigma$ is harmonic, and hence $\dd^\ast$-closed. This confirms the second assertion of the proposition.
\end{prooff}

Making use of Stokes' theorem, it is clear that the result of this proposition leads directly to the localization of the energy integral (\ref{6.1}).
To make the discussion simpler, we assume for the rest of this section that $\Sigma$ is a Riemann surface. Then the condition of proposition \ref{prop6.1} means that $\Sigma^o$ should be equal to $\Sigma$ minus a finite set of points $z_1 , \ldots, z_r$. Let now $D_s^R$ be a small disk of radius $R$ around the point $z_s$, and let $\partial D_s^R$ be the circular boundary of the disk with the conventional boundary orientation. Then Stokes' theorem leads to the following result.
\begin{cor}
Assume that $\Sigma$ is a Riemann surface. Then under the conditions of proposition \ref{prop6.1} the $L^2$-norm of horizontal tangent vectors is given by 
\begin{equation}
|\!|\,(\dA , \dphi)\, |\!|_{L^2}^2 \ =\ \sum_{s=1}^r \ \lim_{R\rightarrow 0} \ \int_{\partial D_s^R} \ (\frac{i}{2} \: k_{ab} \, \bar{\lambda}^a \, \dA^b_{\bar{z}}) \ \dd \bar{z} \ .
\label{6.10}
\end{equation}
\end{cor}
This result implies in particular that the right-hand side of (\ref{6.10}) is a real number, a fact that a priori is not obvious. It is also not difficult to argue that each of the loop integrals in (\ref{6.10}) is finite, not just their sum. This follows for example from the observation that, for any $R_1 > R$,
\[
\lim_{R\rightarrow 0} \ \int_{\partial D_s^R} \ (\: k_{ab} \, \bar{\lambda}^a \, \dA^b_{\bar{z}}\: ) \ \dd \bar{z} \ = \ \int_{D_s^{R_1}} {\mathcal E}(\dA , \dphi) \ {\rm vol}_\Sigma \ + \ \int_{\partial D^{R_1}_s}\ (\: k_{ab} \, \bar{\lambda}^a \, \dA^b_{\bar{z}}\: ) \ \dd \bar{z} \ ,
\]
and the right-hand side is finite.

\section{Examples of localization}

\subsection{Linear non-abelian vortices}

In this section we will apply the localization results obtained above to particular examples of vortex equations. We will always assume that $\Sigma$ has complex dimension one, i.e. that it is a Riemann surface. Our first example is the equations with group $G = U(N)$ acting by left multiplication on the target space $X= M_{N\times N}$ of complex square matrices. Note that in this case $X^o \subset X$ is the submanifold of matrices with non-vanishing determinant. Also, with the usual identifications on linear spaces, the vector field induced by $v\in \g$ and the left action at a point $M \in M_{N\times N}$ can be written as
\begin{equation}
\hat{v}\;|_M \ =\ v\, M \qquad {\rm in} \ \ T_M(M_{N\times N}) \simeq M_{N\times N} \ .
\label{7.1.1}
\end{equation}  
The K\"ahler form on the target is chosen to be
\begin{equation}
\omega_X \ = \ \frac{i}{2} \; {\rm Tr}\: (\dd M \wedge \dd M^\dagger) \ ,
\label{7.1.2}
\end{equation}
so that with the natural choice of $Ad$-invariant inner product on the Lie algebra
\[
k(v_1 , v_2) \ = \ {\rm Tr}\: (v_1^\dagger\,  v_2 ) \ ,
\]
the moment map becomes simply
\begin{equation}
\mu (M) \ = \ - \: \frac{i}{2} \: (M M^\dagger - \tau\: 1) \qquad \ \in \ \ u(N) \ ,
\label{7.1.3}
\end{equation}
with $\tau$ a real constant. This particular example of vortex equations has been thoroughly studied, and it is well-known  that if $(A, \phi)$ is a solution with $\det \phi$ not identically zero, then necessarily $\det \phi$ vanishes only at a finite number of points $z_1 , \ldots , z_r  \in \Sigma$ \cite{Baptista3, Eto-al1}. This means that the first condition of proposition \ref{prop6.1} is satisfied, so that localization is applicable.

Now let $(\dA , \dphi)$ be a horizontal tangent vector at $(A, \phi)$. It follows from the definition of $\lambda^a$ in proposition \ref{prop6.1} that the vectors $\lambda = \lambda^a e_a$ in the complexified Lie algebra are given by
\begin{equation}
\lambda (z) \ = \ \dphi \: \phi^{-1} \qquad \quad {\rm for} \ \ z \neq z_j \ .
\label{7.1.4}
\end{equation} 
Using the first vortex equation, expression (\ref{6.5}) then becomes
\begin{equation}
\dA_{\bar{z}} \ = \ - (\partial_{\bar{z}} \dphi) \: \phi^{-1} \ + \ (\partial_{\bar{z}} \phi) \: \phi^{-1}\, \dphi \ \phi^{-1} \ ,
\label{7.1.5}
\end{equation}
which leads to
\begin{equation}
k_{ab}\, \bar{\lambda}^a \dA^b_{\bar{z}} \ = \ -\: {\rm Tr}\big\{ \; \dphi^\dagger \phi \; \partial_{\bar{z}}[(\phi^\dagger \phi)^{-1} \phi^\dagger \dphi ]\; (\phi^\dagger \phi )^{-1}   \;  \big\} \ .
\label{7.1.6}
\end{equation}
Consider now the gauge-invariant function
\begin{align}
f\ :\ \Sigma \times \MM \ &\longrightarrow \  \{ {\rm hermitian\ matrices}  \}    \label{7.1.7}  \\
(z , \: [A, \phi] )& \ \longmapsto \ \phi^\dagger \phi \: (z) \ .  \nonumber
\end{align}
Using the definition (\ref{1.4}) of the natural complex structure $J_\MM$ on the moduli space, it is easy to check that
\begin{equation}
(\partial_\MM f)\: (\dA , \dphi) \ = \ \frac{1}{2} \big[\: (\dd f)(\dA , \dphi) \ - \ i\: (\dd f)(J_\MM (\dA , \dphi) )\:  \big] \ = \ \phi^\dagger \: \dphi \ ,
\label{7.1.8}
\end{equation}
and similarly
\begin{equation}
(\db_\MM f)\: (\dA , \dphi) \ = \ \dphi^\dagger \: \phi \ .
\label{7.1.9} 
\end{equation} 
Thus if $\xi \in T_{[A, \phi]}\MM$ is a tangent vector represented by $(\dA ,\dphi)$, the combination (\ref{7.1.6}) can be rewritten as
\begin{align}
k_{ab}\, \bar{\lambda}^a \dA^b_{\bar{z}} \ &= \ - {\rm Tr} \: \big[ \: f^{-1} (\db_\MM f)(\xi) \; \partial_{\bar{z}}(f^{-1} \partial_\MM f) (\xi)\: \big] \\
&= \  - {\rm Tr} \: \big[ \: f^{-1} (\partial_{\bar{u}^k} f) \; \partial_{\bar{z}}(f^{-1} \partial_{u^j} f)\:  \big]\ (\dd u^j \otimes \dd\bar{u}^k)(\xi, \xi) \ , 
\label{7.1.10}
\end{align}
where $\{ u^j \}$ is any set of local complex coordinates on the moduli space. Finally, the localization of the previous section leads to the following formulae.
\begin{prop}
For any choice of complex coordinates $\{ u^j\}$ on the vortex moduli space, the K\"ahler form of the $L^2$-metric is locally given by
\begin{align}
\omega_\MM \ &= \ \frac{i}{e^2} \ \left[  \int_\Sigma {\rm Tr} \: \Big\{ h^{z\bar{z}}\: \partial_z \big[ (f^{-1}\,\partial_{\bar{u}^k} f) \; \partial_{\bar{z}}(f^{-1}\, \partial_{u^j} f)\:  \big]  \Big\} \ {\rm vol}_\Sigma \right] \ \ \dd u^j \wedge \dd \bar{u}^k   \label{7.1.11} \\
 &= \  \frac{1}{2\, e^2} \ \left\{ \sum_{s=1}^r\ \lim_{R\rightarrow 0}\  \int_{\partial D_s^R} {\rm Tr} \: \big[\: (f^{-1}\, \partial_{\bar{u}^k} f) \; \partial_{\bar{z}}(f^{-1}\, \partial_{u^j} f)\:  \big]\ \dd\bar{z} \right\} \ \ \dd u^j \wedge \dd \bar{u}^k \ ,
\label{7.1.12}
\end{align}
where $f$ is the gauge invariant matrix function $\phi^\dagger \phi$ on the product $\Sigma \times \MM$, and the $\partial D^R_s$'s are small circles of radius $R$ around the points $z_s \in \Sigma$ where $\det \phi$ vanishes.
\label{prop7.1}
\end{prop}
\begin{rem}
Although the function $f^{-1}$ is singular at the points $z_j$, the very way in which the formulae were obtained shows that the integrand of (\ref{7.1.11}) is a smooth function all over $\Sigma$. It is not, however, the exterior derivative of something everywhere smooth, hence the localization to the circle integrals around the points $z_j$. Note also that in (\ref{7.1.12}) we are not assuming that $\det \phi$ vanishes at $z_j$ with multiplicity 1 -- the formula is valid for any multiplicity.
\end{rem}

$\ $

To take the calculation farther we would like to perform the contour integrals along the small circles $\partial D_s^R$, and this requires some knowledge about the singularity of $f^{-1}$ at the zeros $z = z_s$. For simplicity we assume at this point that we are calculating in the region $\MM_{\rm separated}$ of the moduli space where the vortices do not coincide, which is an open dense subset of $\MM$. In this case $\det \phi$ vanishes at $z=z_s$ with multiplicity one and the kernel $\ker \phi (z_s) \subset \CC^N$ has dimension one. In fact, it can be shown that the moduli space of $d$ separated vortices is parameterized precisely by the different possible choices of distinct points $z_s \in \Sigma$ where $\det \phi$ vanishes and the choices of corresponding kernels $L_s = \ker \phi (z_s) \in \CC {\mathbb P}^{N-1}$, i.e.
\begin{equation}
\MM^d_{\rm separated} \ \simeq \ \{ (z_s , L_s) \in \Sigma \times \CC {\mathbb P}^{N-1}:\ 1\leq s \leq d \ {\rm and}\ z_s \neq z_r  \ {\rm for} \ s\neq r \} \ .
\label{7.1.13}
\end{equation}
This identification is a special case of a result proved in $\cite{Baptista3}$ for compact $\Sigma$ and is generally assumed (on strong grounds) to be true also for $\Sigma = \CC$ \cite{Eto-al1}.

As a first step in the residue calculation we will deal initially with the $N=1$ abelian case, thereby reproducing Samols' original result. The generalization to arbitray $N$ will follow.

\subsubsection{$N=1$ abelian case (Samols' result)}

For non-coincident vortices it can be shown that around the zero $z_s$ each vortex solution can be uniquely factorized as
\begin{equation}
\phi \ = \ (z-z_s) \ a(z , \bar{z}) \ ,
\end{equation}
where the function $a$ is smooth and does not vanish at $z=z_s$. Alternatively, in terms of gauge invariant quantities,
\begin{equation}
|\phi|^2 \ = \ |z-z_s|^2 \  |a|^2 \ .
\end{equation} 
Substituting this expression in the contour integrals of (\ref{7.1.12}) it is then easy to compute the residue
\begin{equation}
\lim_{R\rightarrow 0}\ \int_{\partial D_s^R} {\rm Tr} \: \big[\: (f^{-1}\, \partial_{\bar{u}^k} f) \; \partial_{\bar{z}}(f^{-1}\, \partial_{u^j} f)\:  \big]\ \dd\bar{z} \ = \ 2\pi i \ (\partial_{\bar{u}^k} \bar{z}_s) \ \partial_{\bz} \partial_{u^j} \log|a|^2 \ |_{z = z_s} \ .
\end{equation}
Taking as coordinates on the moduli space $\MM_{\rm separated}$ the positions of the vortices, i.e. $u^r = z_r$, the expression for the K\"ahler form becomes simply
\begin{equation}
\omega_\MM \ = \ \frac{i \pi}{ e^2} \ \sum^d_{r, s =1} \  \partial_{\bz} \partial_{z_s} \: \log \Big| \frac{\phi}{z-z_r} \Big|^2       \ |_{z=z_r} \ \ \dd z_s \wedge \dd \bz_r \ ,
\label{7.1.14}
\end{equation}
and this is Samols' result for the metric in terms of the local behaviour of $|\phi|^2$ around its zeros.
\begin{rem}
As described above, the argument of the logarithm in (\ref{7.1.14}) is the smooth and positive function $|a|^2$, so there is no singularity here. The precise expression of the metric given by Samols in \cite{Samols} can be recovered by expanding $\log |a|^2$ as a Taylor series around $z_r$ and expressing (\ref{7.1.14}) in terms of the Taylor coefficients.
\end{rem}

\subsubsection{$N>1$ non-abelian case}

Our aim here is to explain how Proposition \ref{prop7.1} can be used to write down an expression for the vortex K\"ahler metric in terms of complex coordinates on $\MM_{\rm separated}$. Using the identification (\ref{7.1.13}), we will take as coordinates on the moduli space the location $z_s$ of the zeros of $\det \phi$ and a standard set of coordinates on $\CC {\mathbb P}^{N-1}$. For this we will assume that we are on a region of the moduli space where the lines $L_s = \ker \phi (z_s) \subset \CC^N$ can be written as the span of a vector of the form $(w_s^1 , \ldots , w_s^{N-1} , 1)$. The $w_s^i$'s are then the required coordinates on the projective space. Notice that this assumption does not entail any real loss of generality, because such a configuration can always be reached through a transformation of the type $\phi (z) \rightarrow \phi (z) U$, with $U \in  U(N)$, which is both a symmetry of the vortex equations and an isometry of the moduli space.

The first step in the calculation is to note that on $\MM_{\rm separated}$ each vortex solution can be uniquely factorized around the zero $z_s$ as
\begin{equation}
\phi \ = \ A(z, \bz) \ H(z, z_s , L_s) \ ,
\label{7.1.15}
\end{equation}
where the matrix function $A$ is smooth and invertible at $z=z_s$ and $H$ is defined as 
\begin{equation}
H(z, z_s , L_s) \ = \ \begin{bmatrix}
     1    &  \ \    & \      & - w^1_s \\
     \    & \ddots  & \ \    & \vdots \\
      \   &  \ \    & \ 1      & - w_s^{N-1}  \\
     \    & \ \     &   \    &  z -z_s 
 \end{bmatrix} \ ,
\label{7.1.16}
\end{equation}
with all the blank entries equal to zero. This factorization has been widely used since \cite{Eto-al1}, and the results of \cite{Baptista3} justify it at least for compact $\Sigma$. Observe that by construction the matrix $H$ satisfies $\det H = z- z_s$, at the point $z=z_s$ has kernel $L_s$, and depends holomorphically on $z$, $z_s$ and $L_s$. These are the essential properties of $H$, and we could have chosen a different representative in (\ref{7.1.16}) with these same properties. A different choice of representative, however, would entail a different factorization (\ref{7.1.15}) and a less simple form of the end result (\ref{7.1.17}). In terms of gauge invariant quantities, the factorization above becomes
\begin{equation}
\phi^\dagger \phi \ = \ H^\dagger \ (A^\dagger A)\ H \ .
\end{equation}
It is then easy to compute that the argument of (\ref{7.1.12}) becomes
\begin{align}
{\rm Tr} \: \big[\: (f^{-1}\, \partial_{\bar{u}^k} f) \; \partial_{\bar{z}}(f^{-1}\, \partial_{u^j} f)\:  \big] \ =& \ \  {\rm Tr} \: \Big\{ \;  \big[\: (\partial_{\bar{u}^k} H)\, H^{-1} \: \big]^\dagger \ \partial_{u^j} \big[ \: \partial_{\bz} (A^\dagger A) \ (A^\dagger A)^{-1} \: \big] \;  \Big\} \\
&+\  {\rm  terms \ without\  poles} \ .  \nonumber
\end{align}
Choosing the coordinates $u^k$ on the moduli space to be the $z_s$'s and the $w_s^l$'s leads to a particularly simple formula for the residue, because the matrices $(\partial_{u^k} H ) H^{-1}$ then have only one non-zero entry. More precisely, if for all values of $r= 1, \ldots, d$ we name the coordinates
\[
 \left\{ 
\begin{array}{l l}
  u^l_r \ = w_r^l  & \qquad \text{for $1\leq l \leq N-1$}   \\
  u_r^N \ = \ z_r  &  \ \    \\
\end{array} \right.
\]
the final expression for the vortex K\"ahler form becomes simply
\begin{equation}
\omega_\MM \ = \ \frac{i \pi}{ e^2} \ \sum^d_{r, s =1} \  \sum^N_{j, k =1}\ \partial_{u^j_s}\: \big[ \: ( \partial_{\bz} F ) \: F^{-1} \: \big]_{Nk} \ |_{z=u^N_r} \ \ \dd u^j_s \wedge \dd \bar{u}_r^k \ ,
\label{7.1.17}
\end{equation}
where the matrix function $F = A^\dagger A = (H^{-1})^\dagger \phi^\dagger \phi H^{-1}$ is gauge invariant, smooth and invertible around $z = z_r = u_r^N$. This generalizes  Samols' result to these particularly well studied non-abelian vortices.

\subsection{Toric sigma-models}

Our second example generalizes Samols' localization results in a different direction. We consider here the vortex equations with a torus group $G = T^k$ and a toric target $X$. In this case $\dim_\RR G = \dim_\CC X$ and the complexified group $G^\CC = (\CC^\ast)^k$ acts freely and transitively on an open dense subset of the target, so proposition \ref{prop6.1} is applicable. The main example that we have in mind is the case $X= \CC {\mathbb P}^k$, for which the vortex moduli space was described in \cite{Baptista1}. Just as in the classical abelian Higgs model with $X=\CC$, the moduli space is also parametrized by the different possible locations of special points $z_s$ in $\Sigma$, except that for $X= \CC {\mathbb P}^k$ there are $k+1$ distinct types of points, with each type being mapped by $\phi$ to a different submanifold $\CC {\mathbb P}^{k-1}$ inside $\CC {\mathbb P}^k$. (In the classical abelian Higgs model, recall,  there is only one type of special points in $\Sigma$ -- the points that $\phi$ maps to the origin in $\CC$.) So for instance in the simplest case $X= \CC {\mathbb P}^1$, the moduli space is parametrized by the location in $\Sigma$ of two distinct types of points -- the points $z_s^+$ that get mapped to the south pole of $\CC {\mathbb P}^1$, and the points $z_r^-$ such that $\phi (z_r^-)$ is the north pole. Identifying maps to $\CC {\mathbb P}^1$ with meromorphic functions, one can say that $\MM$ is parametrized by the location in $\Sigma$ of the zeros and poles of $\phi$.

Now, since these toric examples are abelian, both the curvature $\ast F_A$ and the moment map $\mu \circ \phi$ are gauge invariant, and so descend to functions on the product $\Sigma \times \MM$ with values on the Lie algebra ${\rm Lie}\ T^k \simeq i \RR^k$. Using the definition (\ref{1.4}) of the complex structure $J_\MM$ on the moduli space, it is easy to check that for a fixed $z \in \Sigma$
\begin{equation}
\partial_\MM (F_A)_{z\bar{z}}^a \ (\dA , \dphi) \ = \ \nabla_z^A \dA^a_{\bar{z}} \ = \ - \frac{e^2}{2}\, (h_\Sigma)_{z\bar{z}}\, (h_X)_{b \bar{c}}\, k^{ca}\, \lambda^b \ ,
\label{7.2.1}
\end{equation}
where the last equality is the complex Gauss' law (\ref{6.6}) and we have written $(h_X)_{b \bar{c}}$ for the invertible hermitian matrix $h_X(\he_b , \he_c )$. Substituting the curvature for the moment map as prescribed by the second vortex equation, we then get
\begin{equation}
\lambda^a \ = \ 2i\, (h_X)^{a \bar{b}}\, \partial_\MM (\mu \circ \phi)_b \ (\dA , \dphi) \ = \ 2 i \,(h_X)^{a \bar{b}} \,\partial_{u^k} (\mu\circ \phi)_b \ \dot{u}^k \ ,
\label{7.2.2}
\end{equation}
where $\{u^k\}$ is any set of complex coordinates on $\MM$. Thus, just as in the example of linear vortices, after using (\ref{6.2}) and (\ref{6.5}) we conclude that the K\"ahler form on the moduli space is given by
\begin{equation}
\omega_\MM \ = \ \frac{4i}{e^2} \; \int_\Sigma \; k_{ab}\, (h_\Sigma)^{z\bar{z}} \ \partial_z \Big\{ (h_X)^{c\bar{a}}\, \partial_{\bar{u}^k} (\mu \circ \phi)_c \; \partial_{\bar{z}} \big[(h_X)^{b\bar{d}}\, \partial_{u^j} (\mu \circ \phi)_d   \big]    \Big\} \ {\rm vol}_\Sigma \ \ \dd u^j \wedge \dd \bar{u}^k \ . \nonumber
\end{equation} 
The integrand in this expression is the total derivative of something that is smooth everywhere on $\Sigma$ except at the points where the matrix $(h_X)_{b \bar{c}} \circ \phi (z)$ is not invertible. Accordingly, the surface integral will localize to a sum of contributions of those singular points. As would be expected, in the $X =\CC {\mathbb P}^k$ example these singular points are exactly the points whose image by $\phi$ lies in one of the $k+1$ natural $\CC {\mathbb P}^{k-1}$'s inside $\CC {\mathbb P}^k$. Thus the local behaviour of $\phi$ around these points completely determines the K\"ahler form $\omega_\MM$.

For instance in the case of the simplest target $X = \CC {\mathbb P}^1$, the moment map $\mu: \CC {\mathbb P}^1  \rightarrow i \RR$ is essentially the height map on the sphere, and a calculation similar to that of section 7.1.1 shows that
\begin{align}  
\omega_{\MM} \ = \  \frac{i\pi}{ e^2} \  \sum_j\  \Big\{  &\sum_{ z_s^+ \in \ \{{\rm zeros\ of\ } \phi \} } \  \partial_{\bz} \partial_{u^j} \: \log \Big| \frac{\phi}{z-z^+_s} \Big|^2   \ |_{z=z^+_s} \  \  \dd u^j \wedge \dd \bz_s^+ \    \nonumber  \\ 
 + \  &\sum_{ z_r^- \in\ \{ {\rm poles\ of\ } \phi \} } \  \partial_{\bz} \partial_{u^j} \: \log \big| \,  (z-z^-_r)\, \phi\, \big|^2   \ |_{z=z^-_r} \ \ \dd u^j \wedge \dd \bz_r^- \;   \Big\} \ ,  \nonumber
\end{align}
where for notational convenience we have called $\{ u^j \}$ the complete set of coordinates $\{z^+_s , z^-_r \}$ on the moduli space. This is the natural generalization of Samols' formula for meromorphic vortices.

\vskip 25pt
\noindent
{\bf Acknowledgements.}
I would like to thank Nuno Rom\~ao for useful discussions about Samols' localization, and Martin Speight for a conversation about the conjectures of section 5.2. I am partially supported by the Netherlands Organisation for Scientific Research (NWO) through the Veni grant 639.031.616.

\section{Appendix}

In very informal terms, the description of the vortex moduli space in theorem \ref{thm4.1} can be understood as follows. Under the assumptions $(i)$ and $(ii)$ of section 4, the usual Hitchin-Kobayashi-type correspondences guarantee that the moduli space $\MM$ is isomorphic to the moduli space of stable solutions of the single equation $\bar{\partial}^A \phi = 0$ modulo complex gauge transformations. (See \cite{Mundet} for the general theory of these correspondences and \cite{Baptista1} for the toric case.) But this remaining equation $\bar{\partial}^A \phi = 0$ just states that $\phi$ should be a holomorphic section of $\oplus_j L_j$, where the holomorphic structure on this vector bundle is induced by the holomorphic structure on $P \rightarrow \Sigma$ determined by the connection $A$. Therefore, up to complex gauge equivalence, giving a pair $(A, \phi)$ that solves this equation is the same as giving a holomorphic structure on $P$ and a holomorphic section of $\oplus_j L_j$. The point now is that the inequivalent holomorphic structures on a $k$-torus bundle over $\Sigma$ are parameterized by the $k$-fold product of the jacobian $J_\Sigma$. Furthermore, after fixing one such holomorphic structure on $P$, i.e. after fixing a point in $\times_k J_\Sigma$, the space of holomorphic sections of each $L_j \rightarrow \Sigma$ is canonically defined and, by Riemann-Roch, has complex dimension $1-g+ d_j$. Letting this point in $\times_k J_\Sigma$ vary we get a  collection of vector spaces -- one vector space for each different point -- and this determines a complex vector bundle $V_j \rightarrow \times_k J_\Sigma$ of rank $1-g+d_j$, as well as the direct sum bundle $V = \oplus_j V_j$. Each point of the total space of $V$ then corresponds to a holomorphic structure on $P$ and a holomorphic section of $\oplus_j L_j$, or in other words to a solution $(A,  \phi)$ of the equation $\bar{\partial}^A \phi = 0$. These solutions, however, are not all complex gauge inequivalent, because a constant gauge transformation acting on $(A, \phi)$ leaves $A$ invariant but shifts the holomorphic section $\phi$, and so identifies different points on the fibres of $V$. Finally, quotienting by these residual constant gauge transformations corresponds to quotienting the fibres of $V$ by the (complex) torus action $(a)$ described in theorem \ref{thm4.1}. The resulting quotient in each fibre is therefore isomorphic to the toric manifold $F_{d_j , \tau}$, and the moduli space $\MM$ becomes the fibre bundle of the theorem.

\end{document}